\documentstyle[12pt]{article}
\oddsidemargin--0mm
\begin{document}
\renewcommand{\theequation}{\thesection.\arabic{equation}}
\newcommand{\la}{\langle}
\newcommand{\ra}{\rangle}
\newcommand{\pl}{\partial}
\newcommand{\be}{\begin{equation}}
\newcommand{\ee}{\end{equation}}
\newcommand{\ba}{\begin{eqnarray}}
\newcommand{\ea}{\end{eqnarray}}
\def\R{\relax{\rm I\kern-.18em R}}
\def\1{\relax{\rm 1\kern-.27em I}}
\newcommand{\Z}{Z\!\!\! Z}
\newcommand{\ph}{PS_{ph}}
\begin{titlepage}

\vskip 1cm
\begin{center}
{\huge 2D Yang-Mills Theories, Gauge Orbit Spaces and
the Path Integral Quantization}$^{^{^{*}}}$

\vskip 1cm
{\large\bf Sergey V. SHABANOV}$ ^{ ^{ ^{**}}}$

\vskip 1cm
{\em Service de Physique Theorique de Saclay,\\
CEA-Saclay, Gif-sur-Yvette Cedex, F-91191,\\
 France}
\end{center}

\begin{abstract}
The role of a physical phase space structure in a classical and quantum
dynamics of gauge theories is emphasized. In particular, the gauge
orbit space of Yang-Mills theories on a cylindrical spacetime (space
is compactified to a circle) is shown to be the Weyl cell for a
semisimple compact gauge group, while the physical phase space
coincides with the quotient $\R^{2r}/W_A$, $r$ a rank of a gauge
group, $W_A$ the affine Weyl group. The transition
amplitude between two points of the gauge orbit space (between two
Wilson loops) is represented via a Hamiltonian path integral over the
physical phase space and explicitly calculated.
The path integral formula appears to be
modified by including trajectories reflected from the boundary of
the physical configuration space (of the Weyl cell)
into the sum over pathes.

The Gribov problem of gauge fixing ambiguities is considered and
its solution is proposed in the framework of the path integral
modified. Artifacts of gauge fixing are qualitatively analyzed
with a simple mechanical example. A relation between a gauge-invariant
description and a gauge fixing procedure is established.

\end{abstract}

\vskip 2cm

\vskip 1.5cm
\noindent
\underline{\hspace*{8cm}}

\noindent
$^*$Works is supported by an MRT grant of the government of France

\noindent
$ ^{**}$On leave from: {\em Laboratory of Theoretical Physics,
Joint Institute for Nuclear Research, P.O.Box 79, Moscow, Russia}

\noindent
e-mail address: {\bf shabanov@amoco.saclay.cea.fr}
\end{titlepage}

\section{Introduction}
\setcounter{equation}0

This paper is devoted to an analysis of quantum dynamics of gauge theories
on a gauge orbit space.
Its main aim is to establish the path integral
representation of the transition amplitude on the gauge orbit space which
results from the Dirac's operator method of quantizing first-class
constrained systems \cite{dir}.

The whole configuration space of any gauge theory is splinted into a set
of gauge group orbits so that gauge transformations generated by
constraints are shifts along those orbits. Any motion of a system along a
gauge orbit is physically irrelevant and, therefore, only transitions
between distinct gauge orbits carry physical information. So, dynamics
of physical degrees of freedom occurs in the space of gauge orbits which
serves as a physical configuration space.

In Sec.2, we show
with trivial examples of a particle in a box, on a
circles and on a line that
a configuration space structure affects quantum dynamics.
Then we give a definition of the physical phase space in gauge
theories. In Sec.3, we analyze a physical configuration
(and phase) space structure in $2D$ Yang-Mills theories ($QCD_2$) with a
semisimple compact gauge group and show that it differs from an
Euclidean space. Due to a solvability of the model \cite{mig}, \cite{raj},
it is intensively investigated \cite{2d} and regarded as a good toy
model for verifying various ideas and methods proposed for 4D gauge
theories.

Recently, we have proposed a
path integral formula for a transition amplitude on the gauge orbit space
for $4D$ Yang-Mills theories \cite{gr}. Because of the Gribov's
ambiguity \cite{gri}, \cite{sin}, the gauge orbit space of Yang-Mills
potentials cannot be uniquely parametrized by potentials satisfying
a gauge condition chosen.
A gauge condition surface in the total configuration space contains
gauge-equivalent configurations. The gauge orbit space turns out to
be isomorphic only to a certain domain on it which is called a modular
domain \cite{zw}. It was shown in \cite{gr} that the Dirac quantization
method leads to a modified path integral formula. One should include
contributions of trajectories reflected from the modular domain
boundary into the Feynman sum over paths in order to obtain a gauge
invariant transition amplitude.
In Sec. 5, we apply this recipe
to the two dimensional case and obtain an exact quantum amplitude for a
transition between two gauge orbits \cite{pisa}.

When describing quantum dynamics on the gauge orbit space, one needs to
introduce coordinates or a certain set of parameters which span the orbit
space. A choice of the orbit space parametrization is not unique and
implies, actually, fixing a gauge. Some gauge fixing artifacts which might
occur through an inappropriate gauge condition choice are discussed in
Sec. 4.

For the Yang-Mills theory on a cylindrical spacetime, physical degrees of
freedom can be described by spatially homogeneous connections belonging to
the Cartan subalgebra, i.e. the Cartan subalgebra is chosen to be a space
of physical coordinates. However, this space is over complete in a sense
that there are configurations in it which correspond to the same gauge
orbit. The latter does not lead to decreasing a number of physical degrees
of freedom, but it does
reduce a "volume" of the physical configuration space.
It turns out that physically distinct configurations constitute a compact
domain in the Cartan subalgebra, namely, the Weyl cell \cite{pisa}.

The path integral formula resulting from Dirac's operator method
of quantizing first-class constrained system does not contain an
integration over the Weyl cell $K^+_W$. A correct transition amplitude is
given by a sum over all trajectories reflected from the Weyl cell
boundary, which is technically equivalent to carrying  out functional
integration over the Cartan subalgebra $H$ with a sequent symmetrization
the result with respect to the affine Weyl group $W_A$ \cite{pisa},
$K^+_W=H/W_A$.

The paper also contains a few Appendices where some technical and
mathematical points are explained.

\section{Physical phase space in gauge theories}

There is no doubt that phase-space (PS) geometry is one of the main
characteristics of Hamiltonian systems, and classical as well as quantum
dynamics strongly depend on it. Consider, for example, a free
particle on a line. The Hamiltonian reads $H=p^2/2$. If PS is
assumed to be a plane $\R^2$, then classical trajectories are straight
lines perpendicular to the momentum axis in PS and outgoing to infinity.
In quantum theory we have the Hamiltonian spectrum $E=p^2/2,\ p\in \R$,
and wave functions being plane waves $\psi\sim\exp ipx$.

Let us change
the PS topology by compactification of the configuration space to a
finite size
$L$. One can do it in different ways. We may identify the boundary points
of the configuration space to turn it into a circle ${\bf S}^1$ of length
$L$, then the phase space becomes a cylinder $PS=\R\otimes{\bf S}^1$. Another
way is to install infinite walls at the boundary points $x=0,L$ to prevent
a particle from penetrating outside of the interval, then the phase space
is a strip $\R\otimes (0,L)$. The classical motion becomes periodical. The
system returns to its initial state (a PS point) via time $T=L/p_0$, $p_0$
is a particle momentum, if $PS=\R\otimes{\bf S}^1$. For a particle moving
between two walls the period is equal to $T=2L/p_0$ because a particle
has to reflect from both walls to reach an initial PS point (notice,
each reflection changes a sign of a particle momentum). The quantum theories
are also different. For the cylindrical PS the spectrum and wave functions
read $E_n=2\pi^2n^2/L^2,\ n\in\Z$, and $\psi_n\sim\exp(2\pi inx/L)$,
respectively, while for PS being a strip they are $E_n=\pi^2n^2/(2L^2)$, and
$\psi_n\sim\sin(\pi xn/L),\ n=0,1,2,...$ .

A main feature of gauge theories is the existence of unphysical variables
whose evolution is determined by arbitrary functions of time \cite{dir},
while physical quantities appear to be independent of the gauge
arbitrariness. A non-trivial geometry of the physical PS (denoted
below as $PS_{ph}$) may occurs in gauge theories, even if the total PS
is assumed to be an even-dimensional Euclidean space \cite{lv}-\cite{book}.

Let a system with $N$ degrees of freedom have $M$ independent first-class
constraints \cite{dir} $\sigma_a(q,p)=0,\ a=1,2,...,M$. Let $H=H(q,p)$ be
a Hamiltonian of the system such that its Poisson bracket with the constraints
vanishes on the constraint surface, i.e. $\{H,\sigma_a\}= c_{ab}\sigma_b= 0$.
The latter means that the system never leaves the constraint surface in due
course since $\dot{\sigma}_a=\{\sigma_a,H\}=0$. Then our system admits a
generalized dynamical description, namely, on the constraint
surface the time evolutions generated by the Hamiltonians $H$ and $H_E
=H+\lambda_a\sigma_a$, where $\lambda_a$ are arbitrary functions of time,
are physically indistinguishable \cite{dir}. Indeed, the first class
constraints generate gauge transformations of canonical variables
$\delta q= \omega_a\{\sigma_a,q\}$ and $\delta p= \omega_a\{\sigma_a,p\}$,
$\omega_a$ are arbitrary infinitesimal
functions of time\cite{dir}. Any physical quantity $F=F(q,p)$
must be gauge invariant and, therefore,
$\delta F = \omega_a\{\sigma_a,F\} =0$, which leads to the equality
$\dot{F} =\{F,H_E\} = \{F,H\}$ on the surface $\sigma_a=0$.

A general solution of the constraints and equations of motion induced
by $H_E$ depends on $M$ arbitrary functions of time $\lambda_a$, while
any variation of those functions  means no change of a physical state
of the system \cite{dir}. Such a variation is nothing but a gauge
transformation of canonical variables. If we require that any physical
state corresponds just to one point of $\ph$, then we are led to the
following definition of $\ph$. {\em The physical phase space is the
quotient of the constraint surface in the whole PS by gauge transformations
${\cal G}$ generated by all independent first class constraints} $\sigma_a$,
\begin{equation}
\ph = PS\vert_{\sigma_a=0}/{\cal G}\ .
\end{equation}

Let us apply (2.1) to a simple gauge system. Consider a particle moving
in a plane. Let the only constraint be its angular momentum \cite{lv}
$\sigma ={\bf p}T{\bf x}= 0,\ T= -i\tau_2,\ \tau_2$ the Pauli matrix,
{\bf x} and {\bf p} are position and momentum vectors of the particle.
The constraint generates simultaneous rotations of {\bf p} and {\bf x}
since $\delta {\bf p} =\omega\{{\bf p},\sigma\} = \omega T{\bf p}$ and
$\delta {\bf x}=\omega\{{\bf x},\sigma\}= \omega T{\bf x}$ and $T$ is
a generator of SO(2). So, the angular variable turns out to be unphysical,
and only the radial motion  should be of physical interest.
Dynamics of the radial degree of freedom occurs on the constraint surface
${\bf p}\sim {\bf x}$ in the total $PS = \R^4$. By means of  gauge
SO(2)-rotations we can identify any vector {\bf x} with a particular
vector $(x,0),\ {\bf x}^2= x^2$. A momentum vector is simultaneously
reduced to $(p,0)$ by the same gauge transformation since ${\bf p}\sim
{\bf x}$. The continuous gauge arbitrariness is exhausted. Notice that
the relation ${\bf p}\sim {\bf x}$ implies $p\sim x$. However, the
variables $p$ and $x$
are regarded as independent because the proportionality
coefficient is an arbitrary function of time (it can be found only by
solving Hamiltonian equations of motion).

One would assume $\ph$ spanned by $p$ and $x$ to be a plane $\R^2$. But
it is not the case. There remain discrete gauge transformations, the
SO(2)-rotations of the vector $(x,0)$ through the angle $\pi$ which
identify the phase-space points $p,x$ and $-p,-x$ on the phase plane
$\R^2$ (as $p\sim x$, the gauge transformations $x\rightarrow \pm x$
imply the simultaneous change of the momentum sign, $p\rightarrow \pm p$).
Therefore, $\ph$ in the model is a {\em cone} unfoldable into a
half-plane \cite{lv},
\begin{equation}
\ph =\R^2/\Z^2 = cone(\pi)\ .
\end{equation}

A manifestation of the $\ph$ structure (2.2) can be observed in
classical and quantum dynamics. Let the Hamiltonian be
$H=({\bf p}^2 + \omega^2{\bf x}^2)/2 $. After eliminating an unphysical
degree of freedom we have a one-dimensional oscillator with a conic PS.
A classical phase-space trajectory of a harmonic oscillator is an ellipse
with its center at the origin. As the points $p,x$ and $-p,-x$ are
gauge equivalent, the system returns to its initial state (an initial
phase-space point) via time $T_{ph}= \pi/\omega$ rather than the
period $T=2\pi/\omega$, i.e. the physical frequency is doubled \cite{lv}
$\omega_{ph} = 2\pi/T_{ph} = 2\omega$.

In quantum theory, the operator $\hat{\sigma}$ must annihilate
physical states \cite{dir}. Rewriting it via the
destruction and creation
 operators , $\hat{\bf a} = (\sqrt{\omega}
\hat{\bf x} + i\hat{\bf p}/\sqrt{\omega})/\sqrt{2}$
and $\hat{\bf a}^+$,
respectively, we have \cite{weyl}
$\hat{\sigma}|ph\rangle = -i\hat{\bf a}^+T
\hat{\bf a}|ph\rangle = 0$. Therefore $|ph\rangle = \Phi(\hat{\bf a}^+)
|0\rangle,\ \hat{\bf a}|0\rangle \equiv 0$, where $[\hat{\sigma},\hat{\Phi}]
=0$. The latter means that the operator $\hat{\Phi}$ is a function invariant
under the SO(2)-rotations of its argument $\hat{\bf a}^+$. The only
independent invariant which can be built of the vector $\hat{\bf a}^+$ is
its square $\hat{\bf a}^{+2}$; all invariants are functions of it. Thus,
an arbitrary physical Fock state reads
\begin{equation}
|ph\rangle = \sum\limits_{n=0}^{\infty}\Phi_n (\hat{\bf a}^{+2})^n
|0\rangle\ ,
\end{equation}
which yields the physical spectrum $E_n^{ph}= 2\omega(n+1/2)=
\omega_{ph}(n+1/2),\ n=0,1,...,$ of the Hamiltonian $H=\omega
(\hat{\bf a}^+\hat{\bf a} + 1)$. The distance between energy levels
is doubled. This result can be also recovered by quantizing
the conic PS (2.2) in the framework of the
WKB method \cite{lv}, \cite{ufn}.

A further consideration of the ${\ph}$ geometry in mechanical
gauge systems can be found in a review \cite{ufn} or in a
monograph \cite{book}, an application to minisuperspace
cosmology \cite{ze} is given  in \cite{cos}, a relation between the
$\ph$ geometry  and quantum Green function in gauge theories
are discussed in \cite{gr} (see also for a review \cite{lis}).

\section{Phase space structure in Yang-Mills theory on a cylinder}
\setcounter{equation}0
The  $\ph$ definition (2.1) holds for gauge
field theories, i.e. for systems with an infinite number of degrees
of freedom. PS in a field theory is a functional space, which
gives rise to some technical difficulties when applying (2.1).
Nevertheless, for some particular gauge field theories, the quotient
(2.1) can be explicitly calculated. The 2D Yang-Mills theory on
a cylindrical spacetime (space is compactified to a circle ${\bf S}^1$)
\cite{mig}, \cite{raj} exhibits a nice example of that kind. We shall
establish the $\ph$ structure of this theory in the case of an
arbitrary compact semisimple gauge group.

The Lagrangian reads
\be
L=-\frac14\int\limits_0^{2\pi l}dx (F_{\mu\nu},F_{\mu\nu})
\equiv - \frac14\langle F_{\mu\nu},F_{\mu\nu}\rangle\ ,
\ee
where $F_{\mu\nu}= \pl_{\mu} A_{\nu}-\pl_{\nu} A_{\mu} - ig[A_{\mu},
A_{\nu}],\ g$ a coupling constant, $\mu,\nu = 0,1$; the Yang-Mills
potentials $A_\mu$ being elements of a Lie algebra $X$ are periodic
functions of a spatial coordinate, $A_\mu(t,x+2\pi l)=A_{\mu}(t,x)$,
i.e. $l$ is the space radius; the brackets $(,)$ in the integrand (3.1)
stand for the
invariant inner product in $X$. We assume it to be the Killing
form (see Appendix A) and suppose also that the orthonormal basis
(A.5) is introduced in $X$. The spatial coordinate $x\in {\bf S}^1$,
the Lorentz and Lie algebra suffices of $A_\mu$ label degrees of freedom
in the theory.

To go over the Hamiltonian formalism, we determine the canonical momenta
$E_\mu = \delta L/\delta\dot{A}_\mu = F_{0\mu}$, the overdot denotes
the time derivative. The momentum conjugated to $A_0$ vanishes, $E_0=0$,
forming the primary constraints. The canonical Hamiltonian has the
form $H=\langle E_\mu,A_\mu\rangle - L= \langle E_1,E_1\rangle/2 -
\langle A_0,\sigma\rangle$ where $ \sigma= \nabla(A_1)E_1$ with
$\nabla(A_1)= \pl_1- ig[A_1,\ ]$ being the covariant derivative
in the adjoint representation. The primary constraints must be satisfied
during the time evolution \cite{dir}. This yields the secondary
constraints
\be
\dot{E}_0 = \{E_0,H\} =  \pl_1E_1 - ig[A_1,E_1] = \sigma = 0\ ,
\ee
where the standard symplectic structure $\{A_\mu^a(x), E_\nu^b(y)\} =
\delta^{ab}\delta_{\mu\nu}\delta(x-y),\ x,y\in {\bf S}^1$, has been
introduced, the suffices $a,b$ enumerate Lie algebra components.
Since $\{\sigma^a(x),\sigma^b(y)\}$ $= if^{abc}\delta(x-y)\sigma^c(x)$,
$f^{abc}$ are structure constants of $X$, and $\{\sigma_a,H\} =
-f^{abc}A_0^b\sigma^c$,
we conclude that there is no more constraints in the theory, and all
constraints are of the first class \cite{dir}.

The primary and secondary (first-class) constraints are treated as
independent generators of gauge transformations (see Sec.2). It is
readily to see that the primary constraints $E_0^a=0$ generate shifts
of $A_0^a\ (\delta A_0^a(x)= \{A_0^a,\langle\omega_0,E_0\rangle\} =
\omega_0^a(x))$ and leave the phase space variables $E_\mu^a$ and
$A_1^a$ untouched. Therefore the hyperplane $E_0^a=0$ being the
constraint surface in the total PS is a gauge orbit. In accordance
with (2.1) it contributes just a point $E_0=A_0=0$ to $\ph$. Thus,
we can ignore those pure unphysical degrees of freedom and concentrate
our attention just on the remaining variables.

The constraints (3.2) generate following gauge transformations
\be
E_1\rightarrow \Omega E_1\Omega^{-1}=E_1^\Omega\ ,\ \ \ \
A_1\rightarrow \Omega A_1\Omega^{-1} + \frac ig\Omega\pl\Omega^{-1}
=A_1^\Omega\ ;
\ee
here and below $\pl_1\equiv \pl$,
$\Omega = \Omega(x)$ takes its values in a semisimple compact
group $G$ ($X$ is its Lie algebra). The gauge transformed variables
$E_1^\Omega$ and $A_1^\Omega$ must be also periodic functions of $x$.
This results in the periodicity of $\Omega$ modulo the center $Z_G$
of $G$
\be
\Omega(x+2\pi l)=z\Omega(x)\ ,\ \ \ \ \  z\in Z_G\ .
\ee
Indeed, by definition $z$ commutes with any element of $X$ and, therefore,
$E_1^\Omega$ and $A_1^\Omega$ are invariant under the shift $x
\rightarrow x + 2\pi l$.

The relation (3.4) is called a twisted boundary condition \cite{th}.
As has been pointed out in \cite{th}, twisted gauge transformations
(i.e. satisfying (3.4) with $z\neq e$, $e$ a group unit) form distinct
homotopy classes. Therefore they cannot be continuously deformed towards
the identity. On the other hand, gauge transformations generated by the
constraints (3.2) are homotopically trivial because they are built
up by iterating the infinitesimal transformations \cite{jac}
$\delta E_1= \{E_1,\langle\omega,\sigma\rangle\} = ig [E_1,\omega]$
and $\delta A_1= \{A_1, \langle\omega,\sigma\rangle\} =
-\nabla(A)\omega$ with $\omega$ being an $X$-valued periodic function
of $x$. Thus, we are led to the following
conclusion. When determining $\ph$ by means of (2.1), one should
restrict oneself by {\em periodic} (i.e. homotopically trivial)
gauge transformations \cite{pisa}. Such transformations determine
a mapping ${\bf S}^1\rightarrow G$.
Yet we shall see that quantum states annihilated by the operators
of the constraints (i.e. the physical states \cite{dir}) are not
invariant under twisted gauge transformations (see Appendix D)
which confirms that twisted gauge transformations belong
to a homotopically non-trivial class \cite{jac}.

Consider a periodic function $f(x)$ taking its values in $X$.
It is expanded into a Fourier series
\be
f(x)=f_0 +\sum\limits_{n=1}^\infty\left(f_{s,n}\sin\frac{nx}{l}
+f_{c,n}\cos\frac{nx}{l}\right)\ .
\ee
We denote a space of functions (3.5) ${\cal F}$ and its finite dimensional
subspace formed by constant functions ${\cal F}_0$ so that
$A_1= A_{10} + \tilde{A}_1$, where $A_{10}\in {\cal F}_0$ and
$\tilde{A}_1\in {\cal F}\ominus {\cal F}_0$.

Any configuration $A_1$ belongs to the same
gauge orbit as its homogeneous component $A_1=A_{10}$ does. Set
$\Omega^{-1}= P\exp ig\int_0^x\omega dy,\ \omega\in {\cal F}\ominus
{\cal F}_0$,
in (3.3) and require $\pl_1A_1^\Omega =0$ (the Coulomb gauge
$A_1^\Omega\in {\cal F}_0$). Using
the relations $\pl\Omega^{-1}= ig\omega\Omega^{-1}$ and $\pl\Omega
=-ig\Omega\omega$ we derive the equation for $\omega$
\be
\nabla(A_1)\omega = \pl\omega - ig[A_1,\omega] = \pl \tilde{A}_1\ .
\ee
The operator $\nabla(A_1)$ is invertible for $\omega\in {\cal F}\ominus {\cal
F}_0$.
Indeed, a general solution to the homogeneous equation $\nabla(A_1)
\omega =0$ can be written as $\omega^{hom}(x)=W\omega(x_0)W^{-1},\
W= P\exp ig\int_{x_0}^xdy A_1$. For $\omega^{hom}\in {\cal F}\ominus {\cal
F}_0$ there
always exists a point $x_0\in (0,2\pi l)$ such that $\omega^{hom}(x_0)
=0$ because $\int_0^{2\pi l}dx\omega^{hom} = 0$, and, hence, $\omega^{hom}
(x)\equiv 0$ since $W\neq 0$.
So, any configuration $A_1\in {\cal F}$ can be reduced
towards a spatially homogeneous configuration $A_{10}$ by means of
a gauge transformation.

Now we shall prove that the gauge reduction of $A_1$ to $A_{10}$
leads to a simultaneous gauge reduction of the momentum $E_1$
to $E_{10}\in {\cal F}_0$ on the constraint surface (3.2). Substituting
(3.3) into (3.2) and assuming $A_1^\Omega\equiv A_{10}$ we get
$\pl E_1^\Omega - ig[A_{10},E_1^\Omega]= 0$. Putting $E_1^\Omega
=E_{10} + \tilde{E}_1^\Omega,\ \tilde{E}_1^\Omega\in {\cal F}
\ominus {\cal F}_0$
we obtain two equations
\ba
\sigma_0\equiv [A_{10},E_{10}]&=& 0\ ,\\
\pl \tilde{E}_1^\Omega - ig[A_{10},\tilde{E}_{10}^\Omega] &=&0\ .
\ea
Equation (3.8) has only trivial solution $\tilde{E}_1^\Omega = 0$
(see the above discussion of the solution
$\omega^{hom} = 0$ to Eq. (3.6)).

Thus, we are led to a system with $N=\dim X$ degrees of freedom
and the constraint (3.7) which generates homogeneous gauge
transformations of the phase-space variables $A_{10}$ and $E_{10}$
($\pl\Omega\equiv 0$ in (3.3)). This mechanical system has been
studied in \cite{weyl}. The system is shown to have $r=rank\ X$
physical degrees of freedom which can be described by Cartan
subalgebra components of $A_{10}$ and $E_{10}$.

Any element of $X$ can be represented in the form \cite{hel}
$A_{10}=\Omega_Aa\Omega_A^{-1}$, $a$ an element of the Cartan
subalgebra $H$ (see Appendix A), $\Omega_A\in G$. Therefore
configurations $A_{10}$ and $a$ belong to the same gauge
orbit. Moreover, a spatially homogeneous
gauge transformation with $\Omega =
\Omega_A^{-1}$ brings the momentum $E_{10}$ on the
constraint surface (3.7) to the Cartan subalgebra. Indeed,
from (3.7) we derive $[a,\Omega_A^{-1}E_{10}\Omega_A] =0$
and conclude that $p_a=\Omega_A^{-1}E_{10}\Omega_A \in H$
by the definition of $H$. The element $a$ has a stationary
group being the Cartan subgroup of $G$. This means that
not all of the constraints (3.7) are independent. Namely,
there are just $N-r,\ r =\dim H$, independent constraints
amongst (3.7). The continuous gauge arbitrariness is
exhausted in the theory.

One would assume $\ph$ to be $\R^{2r}$ (meaning $H\sim \R^r$)
but this is wrong. There remain discrete gauge transformations
which cannot decrease a number of physical degrees of freedom,
while they do reduce their PS.

As has been pointed out in \cite{weyl}, the mechanical system
with the constraint (3.7) possesses a non-trivial $\ph$ geometry
due to the Weyl group $W$ \cite{hel} being a subgroup of the
gauge group and acting in the reduced PS spanned by $a$ and $p_a$.
Any element of $W$ is a composition of reflections $\hat{s}_\omega$
in hyperplanes orthogonal to simple roots $\omega$, $(a,\omega)=0$
(see Appendix A),
\be
\hat{s}_\omega a = \Omega_\omega a\Omega_\omega^{-1}
=a - \frac{2(a,\omega)}{(\omega,\omega)}\omega
\ ,\ \ \ \ \Omega_\omega\in G\ .
\ee
The Weyl group preserves the root system of $X$ \cite{hel}, p.456. The
group elements $\Omega_\omega$ can be easily found in the orthogonal
basis (A.5),
\be
\Omega_\omega=\Omega_s=\exp\frac{i\pi}{(\omega,\omega)^{1/2}}\ s_\omega
,\ \ \ \  {\rm or}\ \ \ \ \
\Omega_\omega=\Omega_c=\exp\frac{i\pi}{(\omega,\omega)^{1/2}}\ c_\omega\ .
\ee
Using the commutation relations
\be
[s_\omega,c_\omega]= i\omega\ ,\ \ \
[\omega,s_\omega] =i(\omega,\omega)c_\omega\ ,\ \ \ \
[\omega,c_\omega]= -i(\omega,\omega)s_\omega
\ee
one can be convinced that the group elements (3.10) satisfy (3.9). The
existence of two different representations (3.10) of the reflection
operators (3.9) plays the crucial role in gauge dynamics with fermions
\cite{jpa}.

Let $G=SU(2)$. The representation of the Cartan-Weyl basis via
the Pauli matrices is given in Appendix A. Using it we get
$\Omega_s= i\tau_1$ and $\Omega_c= i\tau_2$, $\tau_3$ is the
only basis element of $H_{su(2)}$. Then $(i\tau_1)\tau_3
(-i\tau_1) = -\tau_3 = (i\tau_2)\tau_3 (-i\tau_2) =
\hat{s}_\omega \tau_3$.

Thus, we conclude that the points $\hat{s}p_a,\ \hat{s}a,\
\hat{s}\in W,$ in $\R^{2r}$ are gauge equivalent, i.e. they
belong to the same gauge orbit and, therefore, should be
identified in accordance with (2.1). The Weyl group simply
transitively acts on the set of Weyl chambers \cite{hel},
p.458. Any element of $H$ can be obtained from an element
of the positive Weyl chamber $K^+$ ($a\in K^+$ if $(a,\omega)
> 0,\ \omega$ ranges all simple roots) by a certain transformation
from $W$. In other words, the chamber $K^+$ is the quotient
$H/W$.

In contrast with the mechanical model studied in \cite{weyl},
the Weyl group does not cover the whole admissible discrete
gauge arbitrariness in the 2D Yang-Mills theory. Put $E_1 =
p_a$ and $A_1= a$ in (3.3) and consider such gauge transformations
$\Omega$ which do not transfer $p_a$ and $a$ out of the Cartan subalgebra
and preserve the conditions $\pl a= \pl p_a = 0$. If $\Omega$ does
not belong to the Cartan subgroup $G_H$ and $\pl \Omega =0$, then
it must be an element of $W$ as we have seen above. If $\Omega\in
G_H$ and $\pl_1\Omega=0$, then it is an element of the stationary
group of $a$. Let now $\Omega\in G_H$ and depend on $x$ such
that $\Omega\pl \Omega^{-1}$ is independent of $x$. Obviously,
$\Omega =\Omega_\eta =\exp(ix\eta /l)$ where $\eta\in H$. This transformation
transfers $a$ to $a+a_0\eta,\ a_0= (gl)^{-1}$, and leaves $p_a$ untouched.
The group element $\Omega_\eta$ has to obey the boundary condition (3.4)
with $z=e$ as has been argued above. This yields the equation for $\eta$
\be
\exp(2\pi i\eta)= e\ .
\ee
The set of elements $\eta$ obeying (3.12) is called the unit lattice
in the Cartan subalgebra \cite{hel}, p.305.

Consider the diagram $D(X)$ being a union of a finite number of families
of equispaced hyperplanes in $H$ determined by $(\alpha,a)\in a_0\Z,\
\alpha$ ranges over the root system. Consider then a group $T_e$ of
translations in $H$, $a\rightarrow a+ a_0\eta,$ where $ \eta$ belongs
to the unit lattice. The group $T_e$ leaves the diagram $D(X)$ invariant
\cite{hel}, p.305. The diagram $D(X)$ is also invariant with respect
to  Weyl group transformations. Since $W$ is generated by the
reflections (3.9), it is sufficient to prove the invariance of $D(X)$
under them. We have $(\alpha,\hat{s}_\omega a) = a_0n_\omega$ where
$n_\omega = n - 2k_\omega(\omega,\alpha)/(\omega,\omega)$ is an integer
(see Appendix A),
and $(a,\omega)= k_\omega a_0,\ k_\omega\in \Z$ because $a\in D(X)$. So,
$\hat{s}_\omega D(X)= D(X)$.

Consider the complement $H\ominus D(X)$. It consists of equal
polyhedrons whose walls form the diagram $D(X)$. Each polyhedron
is called a cell. A cell inside of the positive Weyl chamber
$K^+$ such that its closure contains the origin is called
the Weyl cell $K_W^+$. For instance, $D(su(2))$ consists of
points $na_0\omega/(\omega,\omega),\ n\in \Z$ with $\omega$ being
the only positive root of $su(2)$, $(\omega,\omega)= 1/2$ ($\omega
= \tau_3/4$, see Appendix A). A cell of $H_{su(2)}\ominus
D(su(2))$ is an open
interval between two neighbor points of $D(su(2))$. Since
$K^+\sim \R^+$, we conclude that $a\in K^+_W$ if $a_3\in
(0,\sqrt{2}a_0)$ where $a=\sqrt{2}
a_3\omega,\ (a,a)= a_3^2$ in the orthonormal basis.
The translations $a \rightarrow
a+ 2na_0\omega/(\omega,\omega),\ n\in \Z$ form the group
$T_e$, and $W=\Z_2,\ \hat{s}_\omega a = -a$. So, $D(su(2))$
is invariant under translations from $T_e$ and the reflection
from $W$.

For $X=su(3)$ we have three positive roots, $\omega_1,\ \omega_2$
and $\omega_{12}= \omega_1 + \omega_2$ which have the same norms.
The angle between any two neighbor roots is equal to $\pi/3$.
The diagram $D(su(3))$ consists of three families of equispaced
straight lines $(\omega_{1,2,12},a) = a_0n_{1,2,12},\ n\in \Z$ on
the plane $H_{su(3)}\sim \R^2$. The lines are orthogonal
to the roots $\omega_{1,2,12}$, respectively. The complement
$H_{su(3)}\ominus D(su(3))$ is a set of equal-side triangles
covering the plane $H_{su(3)}$. The Weyl cell $K^+_W$ is
the triangle bounded by lines $(\omega_{1,2},a)= 0$ (being
$\pl K^+$) and
$(\omega_{12},a)= a_0$. The group $T_e$ is generated by integral
translations through the vectors $2a_0\alpha/(\alpha,\alpha),\
\alpha$ ranges $\omega_{1,2,12}$, $(\alpha,\alpha)=1/3$ (see
Appendix A).

Since all residual discrete gauge arbitrariness is exhausted
by $T_e$ and $W$, we conclude that there are no gauge equivalent points
in $K^+_W$, i.e. $H\supset CS_{ph}\supseteq K^+_W$ where
$CS_{ph}$ denotes the physical configuration space. It can be
defined analogously to (2.1), $CS_{ph}= CS/{\cal G}$
if gauge transformations generated by constraints do not mix
generalized coordinates spanning the total configuration space $CS$ and
generalized momenta. In the
case of the 2D Yang-Mills theory, $CP_{ph} = [A_1]/{\cal G}$
where ${\cal G}$ is composed of the homotopically trivial
transformations (3.3). Actually, we shall prove the
equality $CS_{ph}=K_W^+$.

Let $W_A$ denote the group of linear transformations of $H$
generated by the reflections in all the hyperplanes in the
diagram $D(X)$. This group is called the affine Weyl group
\cite{hel}, p.314. $W_A$ preserves $D(X)$ and, hence,
\be
K^+_W=H/W_A\ ,
\ee
i.e. the Weyl cell is a quotient of the Cartan subalgebra by
the affine Weyl group. Consider a group $T_r$ of translations
$a\rightarrow a + 2a_0\sum_{\alpha >0}n_\alpha\alpha/(\alpha,
\alpha),\ n_\alpha\in \Z$. Then $W_A$ is semidirect product
of $T_r$ and $W$ \cite{hel}, p.315. The fact that $CS_{ph}$
coincides with $K^+_W$ follows from the equality \cite{hel},
p.317,
\be
\exp\frac{4\pi i\alpha}{(\alpha,\alpha)} = e\ ,
\ee
i.e. $T_e\supseteq T_r$.

Thus, the residual discrete gauge transformations form the
affine Weyl group. The physical phase space is the quotient
\cite{pisa}
\be
\ph = \R^{2r}/W_A\ ,
\ee
where the action of $W_A$ on $H\otimes H\sim \R^{2r}$ is
determined by all possible compositions of the following transformations
\ba
\hat{s}_{\alpha,n}p_a&=&\hat{s}_\alpha p_a =
p_a -\frac{2(\alpha,p_a)}{\alpha,\alpha)}\alpha\ ,\\
\hat{s}_{\alpha,n}a &=&\hat{s}_\alpha a + \frac{2n_\alpha a_0}
{(\alpha,\alpha)}\alpha\ ,
\ea
where the element $\hat{s}_{\alpha,n}\in W_A$ acts on $a$ as a
reflection in the hyperplane $(\alpha,a)= n_\alpha a_0,\ n_\alpha
\in \Z$, and $\alpha$ is any root.

To illustrate the formula (3.15), let us first construct $\ph$ for
the simplest case $X=su(2)$. We have $r=1, \ W=\Z_2,\ (\omega,\omega)
=1/2$. The group $T_r=T_e$ acts on the phase plane $\R^2$ spanned
by the coordinates $p_3,a_3$ (we have introduced the orthonormal
basis in $H_{su(2)}$; see the discussion of $D(su(2))$ above) as
$p_3,a_3\rightarrow p_3, a_3+ 2\sqrt{2}na_0$. So, $\R^2/T_r$ is
a cylinder or the strip $p_3\in \R,\ a_3\in (-\sqrt{2}a_0,
\sqrt{2}a_0)$ with the identified boundary lines $a_3=\pm\sqrt{2}a_0$.
On this strip one should stick together the points $p_3,a_3$ and
$-p_3,-a_3$ connected by the reflection from the Weyl group.
This converts the cylinder into a half-cylinder ended by two
conic horns at the points $p_3=0, a_3=0,\sqrt{2}a_0$. In
neighborhoods of these points $\ph$ looks locally like
$cone(\pi)$ (cf. (2.2)) because $W_A$ acts as the $\Z_2$-reflections
(3.16) and (3.17) with $\alpha = \omega$ and $n=0,1$ near $a_3= 0,
\sqrt{2}a_0$, respectively.

For groups of rank 2, all conic (singular) points of $\ph$ are
concentrated on a triangle being the boundary $\pl K_W^+$ of the
Weyl cell (if $X=su(3)$, $\pl K_W^+$ is an equal-side triangle
with side length $\sqrt{3}a_0$ in the orthonormal basis
defined in Appendix A). Let us  introduce local symplectic
coordinates $p_a^\bot,\ a^\bot$ and $p_a^\| ,\ a^\|$ in a
neighborhood of a point of $\pl K_W^+$ (except the triangle
vertices) which vary along lines perpendicular and parallel
to $\pl K_W^+$, respectively. The $W_A$-reflection in
the wall of $\pl K^+_W$ going through this
neighborhood leaves $p_a^\|,\ a^\|$ untouched, while it
changes the sign of the other symplectic pair, $p_a^\bot,
a^\bot\rightarrow -p_a^\bot,-a^\bot$. Therefore $\ph$
locally coincides with $\R^2\otimes cone(\pi)$. At the
triangle vertices, two conic singularities going along
two triangle edges stick together. If those edges are
perpendicular, $\ph$ is locally $cone(\pi)\otimes
cone(\pi)$. If not, $\ph$ is a $4D-hypercone$. The point
of the $4D-hypercone$ is ``sharper'' than the point of
$cone(\pi)\otimes cone(\pi)$, meaning that the $4D-hypercone$
can be always put inside of $cone(\pi)\otimes cone(\pi)$.
Obviously, a less angle between the triangle edges
corresponds to a ``sharper'' hypercone.

A generalization of this pattern of singular points in $\ph$ to
gauge groups of an arbitrary rank is trivial. The Weyl cell
is an $rD$-polyhedron. $\ph$ at the polyhedron vertices has
the most singular local $2rD-hypercone$ structure. On the
polyhedron edges it is locally viewed as $\R^2\otimes
2(r-1)D-hypercone$. Then on the polyhedron faces, being
polygons, the local $\ph$ structure is $\R^4\otimes
2(r-2)D-hypercone$, etc.

\section{Artifacts of gauge fixing and the Gribov problem}
\setcounter{equation}0

The definition (2.1) of $\ph$ is independent of choosing symplectic
coordinates and explicitly gauge-invariant. However, upon a dynamical
description (quantum or classical) of a constrained system, we need
to introduce coordinates on $\ph$, which means fixing a gauge or
choosing a $\ph$ parametrization. This parametrization is
usually motivated
by physical reasons. If we deal with gauge fields, one may describe
physical degrees of freedom by transverse components ${\bf A}^\bot$
of the vector potential and their canonically conjugated momenta
${\bf E}^\bot$, i.e. the Coulomb gauge
$\mbox{\boldmath$ \pl$}{\bf A}=0$ is imposed
to remove unphysical degrees of freedom. This choice comes from
by our experience in QED where two independent polarizations of
a photon are naturally described by the transverse vector-potential.
Apparently, for QED $\ph \sim [{\bf A}^\bot]\otimes [{\bf E}^\bot]$ where
$[{\bf A}^\bot]$ implies the functional space of all configurations
${\bf A}^\bot$.

 Transverse fields cannot serve as good
variables parametrizing $\ph$ in the non-Abelian case
because there are gauge-equivalent
configurations in $[{\bf A}^\bot]$, Gribov's copies \cite{gri}.
Moreover, this gauge fixing ambiguity always arises and has a geometric
nature \cite{sin} related to topological properties of
the gauge orbit space and cannot
be avoided if gauge potentials are assumed
to vanish at the spatial infinity.
So, one should develop a formalism  taking into account
a true geometric structure of $\ph$ in a quantum dynamical description
\cite{gr}.

In the 2D Yang-Mills theory considered in Sec.3, spatially homogeneous
Cartan subalgebra components of  the vector potential $A=a$ and field
strength $E=p_a$ can be regarded as symplectic coordinates on $\ph$.
In fact, this implies the Coulomb gauge condition $\pl A=0$ which is
not complete in this case because there are
some unphysical degrees of freedom
left. They are removed by imposing the additional gauge condition
$(e_{\pm\alpha}, A)=0$, i.e. $A\in H$. Gribov's copies of a
configuration $A=a \in [a]=H \sim \R^r$ are obtained by applying
elements of the affine Weyl group $W_A$ to $a$. The modular domain (see
Sec.1) obviously coincides with the Weyl cell. If $a$ belongs to the
modular domain boundary $\pl K^+_W$, the residual gauge arbitrariness
contains even continuous transformations. However, such configurations
form a set of zero measure in $[a]=H$ and play no role in quantum
dynamics.

Gribov's copies themselves do not have much physical meaning because
they strongly depend on a concrete choice of a gauge fixing condition
that is rather arbitrary.
To illustrate this, let us return back to a simple mechanical gauge model
of Sec.2. The unitary gauge $x_2= 0$ is most convenient to describe
the physical configuration space being a space of concentric circles.
Suppose for a moment that we do not know the structure of the gauge
orbit space. Then all gauge conditions have to be treated on equal
footing.

Any gauge condition $F({\bf x})=0$ determines a curve on a plane
$\R^2$ over which a physical variable ranges. The curve $F({\bf x})
=0$ must cross each orbit at least once because a gauge choice is
nothing but a parametrization of the gauge orbit space. In the
model under consideration, this yields that the curve has to go
through the origin to infinity. Let us introduce a parametrization
of the gauge condition curve
\be
{\bf x} = {\bf x}(u) = {\bf f}(u)\ ,\ \ \ \ \ u\in \R\ ,
\ee
where ${\bf f}(0) = 0$ and $|{\bf f}|\rightarrow \infty$ as
$u\rightarrow \infty$ so that $u$ serves as a physical variable.
If $f_2=0$ and $f_1=u$, we recover the unitary gauge considered above.

Now we can easily see that an inappropriate choice of ${\bf f}$ might
make a dynamical description  very complicated. Let points ${\bf x}$ and
${\bf x}_s$ belong to the same gauge orbit, then ${\bf x}_s= \Omega_s
{\bf x},\ \Omega_s\in SO(2)$. Suppose the curve (4.1) intersects a
gauge orbit at points ${\bf x}={\bf f}(u)$ and ${\bf x}_s = {\bf f}(u_s)$.
We have also $u_s= u_s(u)$ because ${\bf f}(u_s)= \Omega_s {\bf f }(u)$.
If the structure of gauge orbits is supposed to be unknown, the function
$u_s(u)$ can be found  by solving the following equations
\ba
F(\Omega_s{\bf f})&=&0\ ,\\
\Omega_s(u){\bf f}(u) &=& {\bf f}(u_s(u))\ .
\ea
Solutions of (4.2) (the trivial solution, $\Omega_s =1$, always exists by
the definition of ${\bf f}$)
form a set $S_F$ of discrete residual gauge transformations. Eq. (4.3)
determines an induced action of $S_F$ on the physical variable $u$ (a
representation of $S_F$ in the space of $u\in \R$). The set $S_F$ is not a
group
because for an arbitrary $F$ a composition $\Omega_{s}\Omega_{s'}$ of
two elements from $S_F$ might not belong to $S_F$
(if it does not satisfy (4.2)), while for each $\Omega_s$
there exists $\Omega_s^{-1}$ such that $\Omega_s^{-1}\Omega_s = 1$.

Suppose we have two different solutions $\Omega _s$ and $\Omega _{s'}$
to the system (4.2)-(4.3). Then, in general case, the composition $\Omega
_s\Omega _{s'}$ is not a solution to (4.2), i.e. $F(\Omega _s\Omega _{s'}
{\bf f}(u))=F(\Omega _sf(u_{s'}))\ne 0$ because $u_{s'}\ne u$ whereas
$F(\Omega _sf(u))=0$. The functions $u_s(u)$ determined by (4.3) do not
have a unique analytical continuation to the whole covering space $u\in \R$,
otherwise the composition $u_s\circ u_{s'}=u_{ss'}(u)$ would be uniquely
defined and, hence, one could always find an element $\Omega _{ss'}=
\Omega _s\Omega _{s'}$ being a solution to (4.2), which is not the case.
Moreover, a number of elements in $S_F$ can depend on $u$.

To exhibit these artifacts of gauge fixing, we consider a concrete choice
of ${\bf f}$. A general analysis can be found in \cite{lis},\cite{book}.
Set $f_1= -u_0,\ f_2=-\gamma(2u_0 + u)$ for $u< - u_0$ and $f_1= u,\
f_2=\gamma u$ for $u> -u_0$ where
$\gamma $ and $u_0$ are positive constants. The
curve (4.1) touches circles (gauge orbits) of radii $r= u_0$ and
$r=u_0\gamma_0,\ \gamma_0=\sqrt{1+\gamma^2}$. It intersects twice all
circles with radii $r< u_0$ and $r> u_0\gamma_0$, whereas any circle
with a radius from the interval $r\in (u_0,u_0\gamma_0)$ has four
common
points with the gauge condition curve. Therefore, $S_F$ has one
nontrivial element for $u\in \R_1\cup\R_3,\ \R_1= (-u_0/\gamma_0,
u_0/\gamma_0),\ \R_3= (-\infty,-3u_0)\cup (u_0,\infty)$ and three
nontrivial elements for $u\in \R_2= (-3u_0,-u_0/\gamma_0)\cup
(u_0/\gamma_0, u_0)$.
Since points ${\bf f}(u_s)$ and ${\bf f}(u)$ belong to the same
circle (gauge orbit), the functions $u_s$
have to obey the following equation
\be
{\bf f}^2(u_s)= {\bf f}^2(u)\ .
\ee
Denoting $S_F = S_\alpha$ for $u\in \R_\alpha,\
\alpha = 1,2,3$, we have $S_1=\Z_2,\ u_s(u)= -u;\ S_2$ is determined by
the following mappings of the interval $K_2 = (u_0/\gamma_0,u_0)$
\ba
u_{s_1}(u) &=& -u\ ,\ \ \ \ \ \ \ \ \ \ \ \ \ \ \ \ \ \ \ \ \ \ \ \ \ \
\ \ \ \ \ \ \ \
u_{s_1}:\ \ K_2\rightarrow (-u_0,-u_0/\gamma_0)\ ;\\
u_{s_2}(u) &=& -2u_0 + \gamma_0(u^2 - u_0^2/\gamma _0^2)^{1/2}/\gamma\ ,\ \
u_{s_2}:\ \ K_2\rightarrow (-u_0,-2u_0)\ ;\\
u_{s_3}(u)&=&-2u_0 - \gamma_0(u^2 - u_0^2/\gamma _0^2)^{1/2}/\gamma\ ,\ \
u_{s_3}:\ \ K_2\rightarrow (-2u_0,-3u_0)\ ;
\ea
and for $S_3$ we get
\be
u_s(u) = - 2u_0 - \gamma_0(u^2 -u_0^2/\gamma _0^2)^{1/2}/\gamma \ :
\ \ \ (u_0,\infty)\rightarrow (-3u_0, -\infty)\ .
\ee
The functions (4.6-7) do not have a unique analytical continuation
to the whole domain $\R_2$ and, hence, their composition is ill-defined.
The mappings (4.5-7) do not form a group. Since they realize a representation
of $S_\alpha$, $S_\alpha$ is not a group.

The physical configuration space is, obviously, isomorphic to
 $K= \cup K_\alpha,\
K_\alpha= \R_\alpha/S_\alpha$, i.e. $K_\alpha$ is a fundamental domain
of $\R_\alpha$ with respect to the action of $S_F=S_\alpha$ in $\R_\alpha$,
$\R_\alpha = \cup \hat{s}K_\alpha,\ \hat{s}$ ranges over $S_\alpha$.
Upon solving (4.4) (or (4.2-3)) we have to choose a particular interval
as the fundamental domain. We have put $K_2 = (u_0/\gamma_0,u_0)$ in
(4.5-7). Another choice  would lead to another form of the functions
$u_s$ (to another representation of $S_F$ in $\R_2$). Setting, for
example, $K_2= (-2u_0,-u_0)$ we obtain from (4.4)
\ba
u_{s_1}(u)&=& - 4u_0-u\ ,\ \ \ \  \ \ \ \ \ \ \ \ \ \ \ \ \ \ \ \ \ \ \ \ \
u_{s_1}:\ \ K_2\rightarrow (-3u_0,-2u_0)\ ;\\
u_{s_2}(u)&=& -(u_0^2+\gamma^2(2u_0 +u)^2)^{1/2}/\gamma_0\ ,\ \
u_{s_2}:\ \ \ K_2\rightarrow (-u_0,-u_0/\gamma_0)\ ;\\
u_{s_3}(u)&=& (u_0^2 +\gamma^2(2u_0 +u)^2)^{1/2}/\gamma_0\ ,\ \ \ \!
\ \ u_{s_3}:\ \ K_2\rightarrow (u_0/\gamma_0,u_0)\ .
\ea

To find group elements $\Omega_s(u)$ corresponding to $u_s(u)$, one
should solve Eq.(4.3). Setting $\Omega_s =\exp(-T\omega_s)$ and
substituting (4.5-7) into (4.3), we find
\ba
\omega_{s_1}(u) &=& \pi\ ;\\
\omega_{s_2}(u) &=& \frac{3\pi}{2} - \sin^{-1}\left(
\frac{u_0}{\gamma_0u}\right) - \tan^{-1}\gamma\ ;\\
\omega_{s_3}(u) &=& \frac{\pi}{2} + \sin^{-1}\left(
\frac{u_0}{\gamma_0u}\right) - \tan^{-1}\gamma\ ,
\ea
where $u\in K_2=(u_0/\gamma_0,u_0)$. Elements of $S_{1,3}$ are
obtained analogously. It is readily seen that $\Omega_{s_1}
\Omega_{s_2}\neq \Omega_{s_3}$, etc., i.e. the elements $\Omega_s$
do not form a group. An alternative choice of $K_2$ results in a
modification of the functions (4.12-14).

One would assume that all complications of the $CS_{ph}$ structure,
$CS_{ph}\sim K$, found above are caused by using
{\em non-invariant} variables
for describing physical degrees of freedom. Indeed, we have fixed
a ``crazy'' gauge $F({\bf x})=0$ and gained a complicated set of
residual gauge transformations (Gribov's problem).
However, one can easily turn the
variable $u$ into the {\em gauge-invariant} one by means of a special
canonical transformation. The set $S_F$ will appear again due to
topological properties of such a canonical transformation rather than
due to gauge fixing ambiguities. It will determine the phase space
structure of gauge-invariant canonical variables.

Consider the following canonical transformation of ${\bf x}$ and
${\bf p}$ \cite{gr},\cite{lis}
\ba
{\bf x}&=& \exp(T\theta){\bf f}(u)\ ;\\
p_\theta &=& {\bf p}T{\bf x}= \sigma\ ,\ \ \
p_u = \frac12({\bf p},{\bf x})\frac{d}{du}\ln {\bf x}^2\ ,
\ea
where $\{\theta,p_\theta\}=\{u,p_u\} = 1$ (if $\{x_i,p_j\}=
\delta_{ij}$) all other Poisson brackets vanish. Equality (4.15)
represents a generalization of the polar coordinates ($f_1=u,\ f_2=0$).
Since $p_\theta$ coincides with the constraint, we conclude that $\theta$
is the unphysical variable in the model; $\sigma =p_\theta$ generates
its shifts, whereas $\{\sigma, u\}=\{\sigma ,p_u\}=
0$ and, hence, $u$ and $p_u$ are {\em gauge-invariant}.
Using the decomposition
\be
{\bf p} = p_\theta\frac{T{\bf x}}{{\bf x}^2} + p_u\frac{{\bf x}}{\mu(u)}\ ,
\ee
where $\mu (u)=(d{\bf f}/du,{\bf f})$, and the constraints $p_\theta =0$
we derive the physical Hamiltonian
\be
H_{ph}= \left(\frac12{\bf p}^2 +V({\bf x}^2)\right)\vert_{p_\theta=0}=
\frac12\frac{{\bf f}^2(u)}{\mu^2(u)}\ p_u^2 +
V({\bf f}^2(u))\ .
\ee
Hamiltonian equations of motion generated by (4.18) provide a gauge-invariant
dynamical description.

Let turn now directly to seeking the hidden set of transformations $S_F$.
As we have pointed out above, dynamics is very sensitive to a phase
space structure. So, to complete our gauge-invariant description, one
should establish a structure of the phase space spanned by $u$ and $p_u$.
Let us forget for a moment about the gauge symmetry and
the constraint $p_\theta=0$ induced by it. Since (4.15) is a change
of variables there should be a one-to-one
correspondence between points ${\bf x}\in \R^2$
and $\theta,\ u$. The latter yields a restriction on admissible values
of $\theta$ and $u$, $\theta\in [0,2\pi)$ and $u\in K\subset \R$. To see
this, we allow the variables $\theta$ and $u$ to have their values on
the whole real axis and consider transformations $\theta,\ u\rightarrow
\theta +\theta_s =\hat{s}\theta,\ u_s=\hat{s} u$ such that
\be
{\bf x}(\hat{s}\theta, \hat{s}u)= {\bf x}(\theta, u)\ ,
\ee
i.e. we assume ${\bf f}(u)$ to be a real analytical function on $\R$.
Points $\hat{s}\theta,\ \hat{s}u $ of the $(u,\theta)$-plane
 are mapped to one
point on the ${\bf x}$-plane. To provide the mapping (4.15) to be
one-to-one (otherwise it is not a change of variables), one should
restrict values of $\theta$ and $u$ by the modular domain $\tilde{K}=
\R^2/\tilde{S}$ where transformations from $\tilde{S}$ are defined by
(4.19). The set $\tilde{S}$ is decomposed into the product $T_e\times
S_F$ where elements of $T_e$ are translations of $\theta$ through the
group manifold period,
\be
T_e\ :\ \ \ \ \ \theta\rightarrow \theta + 2\pi n, \ \ \ \
u\rightarrow u,\ \ \ n\in \Z\ ,
\ee
and $S_F$ coincides with the set of residual gauge transformations
described above. Let $\Omega_s= \exp(-T\omega_s(u))$, then it is
easily to be convinced that transformations
\be
S_F\ :\ \ \ \ \theta\rightarrow \theta + \omega_s(u)\ ,\ \ \ \
u\rightarrow u_s(u)
\ee
satisfy (4.19) by $\theta_s=\omega_s$. So, $\tilde{K}\sim [0,2\pi )
\cup K$. In the case of the polar coordinates, $S_F=\Z_2,\ \omega_s
=\pi$ and $u_s= -u$, hence $K\sim \R_+$ (a positive semiaxis).

Under the transformations (4.20), the canonical momenta (4.16) remain
untouched, while
\be
p_\theta \rightarrow p_\theta\ ,\ \ \ \ \ \ \
p_u\rightarrow \left(\frac{du_s}{du}\right)^{-1}p_u\equiv p_{u_s}
=\hat{s}p_u
\ee
under (4.21). In the new canonical variables, a state ${\bf p,x}$
corresponds to phase-space points $(p_\theta, \hat{s}\theta,\hat{s}p_u,
\hat{s}u)$, $\hat{s}$ runs over $S_F$ (assuming $\theta\in [0, 2\pi)$).
Therefore, configurations of new canonical variables connected
with each other by $S_F$-transformations are not physically distinguishable.

Consider a phase-space plane, where $p_\theta =0$ and $\theta$ has a
fixed value, and states $(p_\theta=0,\theta,\hat{s}p_u, \hat{s}u)$ on
it. These states differ from each other by values of the angular
variable $(p_\theta,\theta, \hat{s}p_u, \hat{s}u)\sim (p_\theta,
\hat{s}^{-1}\theta, p_u, u)$ where $\hat{s}^{-1}\theta = \theta -
\omega_s(u)$. If now we switch on the gauge symmetry, the angular
variable becomes unphysical and, hence, the difference between all those
states disappears. They correspond to the same physical state. Thus,
the transformations $u,p_u\rightarrow u_s,p_{u_s}$ of the phase plane
turn into the gauge ones so that we have to stick together all points
$\hat{s}u,\hat{s}p_u,\ \hat{s}\in S_F$, to obtain $\ph$ spanned by
$u$ and $p_u$.
For the polar coordinates, we obviously get $\ph
=cone(\pi)$. If we do not care whether our gauge-invariant variables naturally
span the gauge orbit space, we can gain a very complicated structure
of $\ph$, which can make a dynamical analysis hopeless.

One should emphasize that in our approach transformations
$\hat{s}\in S_F$ in the $(u,p_u)$-plane cannot be regarded as the ones
generated by the constraint $\sigma=p_\theta $ since $\{\sigma,u\}=
\{\sigma,p_u\}=0$ in contrast with the gauge fixing description
considered above. Physical variables are
chosen so that the set $S_F$ determining their phase space coincides
formally with the set of residual gauge transformations in the gauge
fixing approach.
Thus, one can always construct gauge-invariant variables such that their
configuration (phase) space coincides with a surface determined by a gauge
condition chosen and therefore, all artifacts inherent to an ``inappropriate''
gauge fixing may be emerged
in a gauge-invariant approach.

To demonstrate
qualitatively what kind of artifacts might occur through the
``inappropriate'' parametrizing $\ph$, we compare phase-space trajectories
in the canonical variables $r=|{\bf x}|,\ p_r=({\bf x,p})/r$ and
$u,\ p_u$. They are connected by the canonical transformation $r=r(u)=
|{\bf f}(u)|,\ p_r=rp_u/\mu = p_u(dr/du)^{-1}$. We also assume the function
${\bf f}$ to be differentiable such that $dr/du =0$ only at two points
$u=u_{1,2}$ and $dr/du>0$ as $u< u_2$ and $u>u_1$, while $dr/du < 0$ if
$u\in (u_2,u_1)$. Our assumption means that the curve ${\bf x}={\bf f}(u),\
u\in \R_+$, goes from the origin, crosses the circle $|{\bf x}|= r_1 = r(u_1)$
at ${\bf x}= {\bf f}(u_1')$ and reaches the circle $|{\bf x}| = r_2
=r(u_2)$, touches it at ${\bf x}= {\bf f}(u_2)$ and turns back to the
circle $|{\bf x}| = r_1$ and after touching it at the point ${\bf x}
={\bf f}(u_1)$ tends to infinity, crossing the circle $|{\bf x}| =r_2$ at
${\bf x}={\bf f}(u_2')$.

In a neighborhood of the origin, $PS(p_r,r)$ has the conic structure
as we have shown above. This local structure is preserved upon the
canonical transformation to the variables $u, p_u$ because it is
a smooth and one-to-one mapping of the strip $r\in (0,r_1)$ on $
u\in (0,u_1')$. The same is related to the half-planes $r>r_2$ and
$u>u_2'$. Troubles rise up on the domain $r\in (r_1,r_2)$ where the
inverse function $u=u(r)$ becomes multi-valued; it has three branches
in our particular case. States belonging to the strips $u\in (u_1',u_2),
\ u\in (u_2,u_1)$ and $u\in (u_1,u_2')$ are physically equivalent because
there are transformations from $S_F$ mapping the strips on each other
and leaving points $p_r, r\in (r_1,r_2)$ untouched.

To understand what might happen to phase-space trajectories in the
PS region $u\in (u_1',u_2')$, consider a motion with a constant
momentum $p_r$ and suppose that a particle is outgoing from the
origin $r=0$. On the $(p_u,u)-$plane, the particle motion corresponds
to a point running along a curve going from the origin $u=0$. As soon as
the PS point reaches the line $u=u_1'$, there appear two ``phantom''
PS trajectories outcoming from the point $p_u=0, u=u_1$ (notice,
$p_{u_1}=p_{u_2}\equiv 0$ since $dr/du = 0$ at $u=u_{1,2}$, and
$u_1'$ is $S_F$-equivalent to $u_1$). If $u_{s_1}$ and $u_{s_2}$ map
$(u_1',u_2)$ onto $(u_2,u_1)$ and $(u_1,u_2')$, respectively, such that
$r(u)=r(u_{s_1})= r(u_{s_2}),\ u\in (u_1',u_2)$, then the ``phantom''
trajectories are described by the pairs $\hat{s}_{1,2}p_u, \hat{s}_{1,2}u$
(cf. (4.22)) where $p_u,u$ range the trajectory in the PS region $u\in
(u_1',u_2)$. Since $du_{s_1}/du <0$ and $du_{s_2}/du >0$, the ``phantom''
trajectory $\hat{s}_2p_u,\hat{s}_2u$ goes to the infinity, while
the point $\hat{s}_1p_u, \hat{s}_1u$ runs in the opposite direction.
The points $p_u,u$ and $\hat{s}_1p_u,\hat{s}_1u$ arrive at $p_u=p_{u_2}
=0, u=u_2$ in the same time and annihilate each other, whereas a
``phantom'' particle moving along the branch $\hat{s}_2p_u,\hat{s}_2u$
approaches the line $u=u_2'$. In the next time moment a particle
leaves the interval $r\in (r_1,r_2)$ (or $u\in (u_1',u_2')$).

Such ``branching'' of classical PS trajectories is a pure artifact of
an ``inappropriate'' parametrization of $\ph$ (or, as we have argued
above, of gauge fixing). It has to be removed by gluing all the
``phantom'' trajectories (branches). In so doing, we cannot however
avoid breaking PS trajectories at the singular points $u=u_{1,2}$.
Indeed, consider trajectories approaching the line $u=u_1'$ with
{\em different} momenta $p_u$ from the origin and crossing it.
The element $\hat{s}_1$ maps these trajectories with $u>u_1'$ onto
trajectories outcoming from just {\em one} PS point $p_u=0,u=u_1'$, but
$\hat{s}_1$ does not touch these trajectories with $u<u_1'$ at all. So,
we will gain breaking of PS trajectories at $u=u_1'$ after gluing
points $p_u,u$ and $\hat{s}_1p_u, \hat{s}_1u$. The same occurs
at the line $u=u_2'$ and the singular point $p_u=0, u=u_2$. Notice
also that points $p_u=0,u_1$ and $p_u=0, u_1'$ are
 stuck together as well
as $p_u=0,u_2$ and $p_u=0,u_2'$, i.e. trajectories in the PS domain
$u\in (u_2,u_1)$ change their orientation when gluing.

To construct $PS(p_u,u)$, we take the half-plane $u\geq 0, p_u\in \R$,
cut it along the lines $u=u_{1,2}$ and $u=u_{1,2}'$, remove the strips
$u\in (u_1',u_2)$ and $u\in (u_1,u_2')$. Then we glue the semi-axes
$u=0,p_u>0$ and $u=0,p_u<0$ (the conic structure at the origin is
preserved!). To complete constructing $\ph(p_u,u)$, we have to identify
the points $p_u=0,u=u_{1,2}'$ with $p_u=0, u=u_{1,2}$, respectively,
 so that the phase space looks like a chain of the ``cut'' cone,
the strip $u\in (u_1,u_2)$ and the half-plane $u>u_2'$ coupled just
by a point-like bridges. We cannot identify the lines $u=u_1$ and
$u=u_1'$ (as well as $u=u_2$ and $u=u_2'$) because there is no
PS trajectory in the PS region $u\in (u_1,u_2)$ reaching the
lines $u=u_{1,2}$ with a non-zero momentum $p_u$; all trajectories
inside the strip $u\in (u_1,u_2)$ fall on the singular points
$p_u=0,u=u_{1,2}$ being artificial attractors created by gauge
fixing. These attractors correspond to zeros of the Faddev-Popov
determinant $\mu(u_{1,2})=0$ \cite{lis}.

We conclude this section with a few brief remarks.

1. Our analysis of gauge fixing artifacts can be generalized to
gauge systems with many degrees of freedom \cite{ufn}, \cite{book}
(even to gauge field theories \cite{lis},\cite{gr}). It can be achieved
by treating $u$ and ${\bf f}(u)$ as elements of (in)finite dimensional
Euclidean spaces; $\Omega$ becomes an element of a gauge group generated
by all independent first-class constraints; the condition $F=0$ has to
fix a gauge completely, i.e., it has to remove all unphysical degrees
of freedom.

2. A construction of gauge-invariant symplectic variables (meaning that
they commute with constraints) might turn
out to be not very helpful for analyzing dynamics if the gauge orbit
space is unknown. A phase space of those variables may have artificial
attractors and ``branching'' trajectories, i.e. all artifacts inherent
to the gauge fixing approach.

3. Nevertheless, if the analysis of the gauge orbit space is complicated
(or there is a reason to prefer a particular gauge condition to the others),
one can develop a quantum theory in the curvilinear coordinates (4.15)
or in their generalization to gauge systems with a higher number of
degrees of freedom \cite{lis}, \cite{book}. A summary of this approach
is given in Appendix B.

4. The variables $a$ and $p_a$ we constructed in Sec. 3 look gauge
non-invariant (they are related to the Coulomb gauge). In virtue of a
canonical transformation analogous to (4.15)-(4.16), one can always turn
$a$ and $p_a$ into gauge-invariant symplectic variables. Dynamics in these
invariant variables occurs on the phase space (3.15) (see Sec. 6).

\section{The Hamiltonian path integral and a configuration space topology}
\setcounter{equation}0
In the previous sections we have seen that the $\ph$ structure plays
an important role  in classical mechanics of gauge theories. The path
integral formalism \cite{fey} is a natural bridge between classical and
quantum mechanics because it allows us to formulate a quantum theory in
terms of classical quantities. A transition of a classical system from
an initial state to the final one is described by a phase-space trajectory
connecting two points on PS (the initial and final states) and satisfying
Hamiltonian equations of motion. After quntization, a transition amplitude
is determined by a sum over all trajectories connecting initial and final
configuration space points, i.e. by the path integral \cite{fey}. Therefore
one might expect that such a sum depends on a topological structure of a
space formed by these trajectories.

To elucidate what happens to the path integral (PI) representation of the
transition amplitude upon changing
a topology of PS or CS, we first analyze the problem in the framework of
the operator formalism. A reason for this is rather obvious.  The operator
and PI approaches are two languages for describing the same theory. If
we have a solution of a problem in one of them, one can always "translate"
it to the other language  by means of some basic rules \cite{fey},\cite{kla}.
In the operator formalism, the transition amplitude obeys the Schroedinger
equation, while topological properties of CS are taken into account by
imposing some boundary conditions. Thus, the problem is reduced to solving
a standard mathematical task.

Consider, for example, a free one-dimentional motion. The transition
amplitude $\ \ \ $ $U_t(x,x')=\langle x|\exp(-it\hat{H}|x'\rangle,\ \hat{H} =
-\pl_x^2/2$ being the Hamiltonian, satisfies the Schroe\-din\-ger equation
\be
i\pl_tU(x,x')=-\frac12\pl_x^2U_t(x,x')
\ee
with the initial condition
\be
U_{t=0}(x,x')=\langle x|x'\rangle =\delta (x-x')\ .
\ee
If $CS=\R$, a solution to (5.1-2) is well-known \cite{fey}
\ba
U_t(x,x')&=&(2\pi it)^{-1/2}\exp \frac{i(x-x')^2}{2t} =\\
&=&\int\limits_{\R^2}\prod\limits_{\tau =0}^t
\left(\frac{dp(\tau)dx(\tau)}{2\pi}\right)
\exp i\int\limits_0^td\tau\left(p\dot{x} -\frac12 p^2\right)\ ,
\ea
where the measure in (5.4) implies a sum over all trajectories
$x(\tau)$ going from $x'=x(0)$ to $x=x(t)$.

Let now CS be compactified
to a circle or to a strip (a particle on a circle or in a box). Then
in addition to Eqs.(5.1-2), the transition amplitude has to obey the
boundary conditions
\ba
U_t^c(x+L,x')&=&U_t^c(x,x'+L)=U_t^c(x,x')\ ,\\
U_t^b(0,x')&=&U_t^b(L,x')=U_t^b(x,0)=U_t^b(x,L)=0\ ,
\ea
with $L$ being a configuration space volume, for a particle
on a circle and in a box, respectively. To obtain a solution
to (5.1-2) and (5.5) or (5.6), one should take the following
linear combinations of (5.3) \cite{q},\cite{kl},\cite{book}
\ba
U_t^{c,b}(x,x')&=&\int\limits_{-\infty}^\infty dx''U_t(x,x'')Q^{c,b}
(x'',x')\ ,\\
Q^b(x'',x')&=&\sum\limits_{n=-\infty}^\infty
\left[\delta(x''-x'-2Ln) - \delta(x''+x'-2Ln)\right]\ ,\\
Q^c(x'',x')&=&\sum\limits_{n=-\infty}^\infty
\delta(x''-x'-Ln)\ ,
\ea
where $x''\in\R$ and $x,x'\in (0,L)$. From the technical point
of view, the kernels (5.8-9)
are analytical continuations of the unit operator kernel (5.2) to
the whole axis. They can be obtained by straightforward summing
the spectral decomposition of the unit operator kernel
$\la x''|x'\ra$ assuming $x''\in \R$.
For example, in the case of a particle
on a circle we have
\ba
\la x''|x'\ra&=&\sum\limits_E\psi_E(x'')\psi_E^*(x')=
L^{-1}\sum\limits_{n=-\infty}^\infty\exp\left(\frac{2\pi in}{L}
(x''-x')\right)=\\
&=&\sum\limits_{n=-\infty}^\infty\frac{2\pi}{L}
\delta\left(\frac{2\pi}{L}(x''-x')-2\pi n\right)=Q^c(x'',x')
\ea
for $x''\in \R$ and $x'\in (0,L)$. The kernels (5.8-9) contain
all information about symmetry properties of wave functions
satisfying either periodic or zero boundary conditions on the
interval $(0,L)$. The representation (5.7) provides keeping
these properties during the time evolution of any state because
the evolution operator kernel (5.7) possesses theses properties
for its left argument $x$ and, hence, its application to any state
$\psi _0(x)$ always
produces a state with the same symmetry properties, $\psi _t(x)=
\int _{0}^{L}dx'U_t^{c,b}(x,x')\psi _0(x')$.

Doing the integral over $x''$ in (5.7) we see that the transition
amplitude in CS with a non-trivial topology is given by a sum
over paths outgoing from not the only initial point.
For a particle on a circle, the sum contains trajectories going
 from $x'+Ln, n\in\Z$ to $x$. Because of the periodicity
of the action, we can interpret such a sum as a sum over all
trajectories with all possible winding numbers $n$. The latter
provides fulfilling the boundary conditions (5.5) or incorporating
the CS topology into the path integral formalism.

In the case of a particle in a box, contributions of trajectories
going from $x'+2Ln$ to $x$ and of those going from $-x'+2Ln$ to $x$
have opposite signs, which results from the boundary conditions (5.6).
The straight trajectories $x'+2Ln\rightarrow x$ can be interpreted as
continuous trajectories connecting $x',x\in(0,L)$ with $2n$ reflections
from the walls at the interval boundaries because they have the same
action. Contributions of the trajectories $-x'+2Ln\rightarrow x$ are
equivalent to contributions of trajectories inside of the box
with an odd number of reflections $2n+1$.

A general analysis of boundary conditions (or CS topology) in the path
integral formalism is given in \cite{book}. A conclusion is similar
to that we have found above -- a topology of CS (or PS) can be taken
into account by including additional "reflected" trajectories into a sum
over paths. For gauge theories with a non-trivial $\ph$, this statement
is proved in Appendix B.

\section{The Hamiltonian PI on the gauge orbit space for 2D Yang-Mills
theory}
\setcounter{equation}0
In this section we show that the idea of including reflected trajectories
into a sum over paths can be successfully applied to 2D Yang-Mills theories
to construct PI on the
gauge orbit space. PI for the 4D-case has been obtained in
\cite{gr},\cite{lis}.
An advantage of considering  2D gauge field theories is that they are
exactly solvable. Therefore, they give nice toy models for verifying
ideas invented for gauge field theories of a general type.

Following the Dirac method we change the canonical variables
$E_1(x)\rightarrow -i\hbar
\delta/\delta A(x)$, $A_1(x)\rightarrow A(x), A(x)\in
{\cal F}$,
by operators
$\footnote{We ignore the pure unphysical degree of freedom $A_0,E_0$
\cite{jac}.}$
and get the quantum theory in the Shroedinger functional representation
\cite{jac}
\ba
\hat{H}\Phi_n[A] = -\frac{\hbar}{2}\langle\frac{\delta}{\delta A},
\frac{\delta}{\delta A}\rangle\Phi_n[A]&=&E_n\Phi_n[A]\ ,\\
\hat{\sigma }\Phi _n[A]=\nabla(A)\frac{\delta}{\delta A}\Phi_n[A]&=&0\ .
\ea
In accordance with the general method proposed in Appendix B,
to solve Eq.(6.2) and to project the Hamiltonian in (6.1) onto
the gauge orbit space, one should introduce
curvilinear coordinates associated with both a gauge transformation
law and a gauge condition chosen \cite{pisa} (cf.(4.15) and (B.4))
\be
A(x)=\Omega (x)a\Omega ^{-1}(x)+\frac{i}{g}\Omega (x)\pl \Omega ^{-1}(x),
\ \ \Omega (x)\in {\cal G}/{\cal G}^0_H,\ \ a\in K^+_W\ ,
\ee
where ${\cal G}$ is a gauge group, ${\cal G}^0_H$ is spatially homogeneous
Cartan subgroup of ${\cal G},\ \ G_f={\cal G}^0_H$. In Sec.3, any
configuration
$A_1(x),\ x\in {\bf S}^1$, is shown to be uniquely represented
in the form (6.3). The condition $a\in K^+_W$ is imposed to provide
a one-to-one correspondence between the "old" and "new" variables. If
we assume $\Omega(x)\in {\cal G}/{\cal G}^0_H$ and $a\in H\sim \R^r$,
then configurations $\Omega(x)\Omega^{-1}_s,\ \hat{s}a,\ \hat{s}$ ranges
the affine Weyl group $W_A$ (see (3.17), (3.10) and (3.9)), are mapped
to the same configuration $A(x)$ by (3.6). That is why admissible values
of $a$ in (6.3) have to be restricted by the Weyl cell $K_W^+$, otherwise
equality (6.3) does not determine a change of variables. Thus, we
parametrize the gauge orbit space $[A_1]/{\cal G}={\cal F}/{\cal G}
\sim K_W^+$ by the variables $a$. We show below (see (6.20)) that the
constraint operator $\hat{\sigma }$ commutes with $a$ and, therefore,
$a$ is a gauge-invariant variable.

For sequential calculations we introduce the following decomposition
of the functional space (3.5)
\be
{\cal F}=\sum\limits_{n=0}^\infty\oplus{\cal F}_n =
\sum\limits_{n=0}^\infty\oplus({\cal F}_n^H\oplus\tilde{\cal F}_n)\ ,
\ee
where ${\cal F}_0$ is a space of constant Lie algebra-valued functions
(the first term in the series (3.5)), ${\cal F}_n, n\neq 0$, is a space
of functions with the fixed period $2\pi l/n$ (a term in the sum (3.5)
with a fixed $n$). Each subspace of ${\cal F}$ is finite-dimensional,
$\dim {\cal F}_0=\dim X,\ \dim {\cal F}_n=2\dim X, n\neq 0$ (we recall
that Lie algebra-valued functions are considered). Functions belonging
to ${\cal F}_n^H$ take their values in the Cartan subalgebra $H$, while
$\tilde{\cal F}_n $ is composed of $2\pi l/n$-periodic functions with
values in $X\ominus H$. All subspaces introduced are orthogonal with
respect to the scalar product $\langle,\rangle=\int_0^{2\pi l}dx(,)$.

The differential $\delta A\in {\cal F}$ can be represented in the form
\be
\delta A =\Omega\left(da -\frac ig\left(\pl\delta w -ig[a,\delta w]
\right)\right)\Omega^{-1} = \Omega\left(da -\frac ig \nabla(a)
\delta w\right)\Omega^{-1}\ ,
\ee
where by the definition of the change of variable $da\in {\cal F}_0^H$ and
$\delta w(x)= i\Omega^{-1}\delta \Omega\in {\cal F}\ominus{\cal F}_0^H$.
Therefore the metric tensor (B.5) reads
\be
\langle\delta A,\delta A\rangle= 2\pi l(da,da) - g^{-2}
\langle\delta w,\nabla^2(a)\delta w\rangle\ .
\ee
Equality (6.6) results from the trivial observation $\langle da,
\nabla(a)\delta w\rangle = -\langle\nabla(a)da, \delta w\rangle = 0$
which is due to $\pl da=0$ and $[da,a]=0$. In the notations of Appendix B,
we have $g_{11}=2\pi l,\ g_{12}=g_{21}=0,\ g_{22}= -g^{-2}\nabla^2(a)$,
where the operator $\nabla(a)$ acts in the space ${\cal F}\ominus
{\cal F}_0^H$. It has no zero mode on this space if $a\in K_W^+$ and,
hence, is invertible.

To obtain the scalar product (B.13), one has to calculate the determinant
$\det g_{AB} =\det g_{11}\det g_{22} = (2\pi l)^r\det(-g^{-2}\nabla^2(a))$.
Consider the orthogonal decomposition
\be
\tilde{\cal F}_n = \sum\limits_{\alpha>0}\oplus{\cal F}_n^\alpha\ ,
\ee
where ${\cal F}_n^\alpha$ contains only $2\pi l/n$-periodic functions
taking their values in the two-di\-men\-sion\-al subspace $X_\alpha\oplus
X_{-\alpha}$ of the Lie algebra $X$ (see Appendix A, (A.1-4)).
The subspaces ${\cal F}^H_n,\ {\cal F}^\alpha _n$ are invariant subspaces of
the operator $\nabla (a)$; it has a block-diagonal form in the in the
decomposition (6.4), (6.7). Indeed, we have $\nabla (a)
=\pl -ig\hat{a}$, where $\hat{a} =[a,\ ]$ is the adjoint operator acting
in $X$ (see Appendix A).
The operator $\pl $ is diagonal in the algebra space,
and its action leaves periods of functions
untouched, i.e. ${\cal F}^{H,\alpha}
_n$ are its invariant spaces.
Obviously, $\hat{a} {\cal F}^H_n=0$ and $\hat{a}
{\cal F}^\alpha _n={\cal F}^\alpha _n$ if $(\alpha ,a)\neq 0$ with accordance
with (A.2). Therefore an action of the operator
$\nabla (a)$ on ${\cal F}\ominus {\cal F}^H_0$
is given by an infinite-dimensional,
block-diagonal matrix. Its blocks have the form
\ba
\nabla ^H_n(a)&\equiv &\nabla (a)\vert _{{\cal F}^H_n}=\pl \vert _{{\cal
F}^H_n}=
\left (\otimes \frac{n}{l}T\right )^r,\ \ n\neq 0,\ \ r=rank X,\\
\nabla ^\alpha _0(a)&\equiv & \nabla (a)\vert _{{\cal F}^\alpha _0}=
-ig\hat{a}\vert _{{\cal F}^\alpha _0}=-ig(a,\alpha )T\\
\nabla ^\alpha _n(a)&\equiv& \nabla (a)\vert _{{\cal F}^\alpha _n}=\1 \otimes
\frac{n}{l}T-ig(a,\alpha )T\otimes \1
\ea
where $T=-i\tau _2,\ \ \tau _2$ the Pauli matrix, $\1$ is the $2\times 2$
 unit matrix. In (6.10) the first components in the tensor products
 correspond to the algebra indices, while the second ones determine the
 action of $\nabla (a)$ on the functional basis $\sin xn/l,\ \ \cos nx/l$.
We have
\ba
\det g_{AB}&=& (2\pi l)^r\prod \limits _{\alpha > 0}
\det (ig^{-1}\nabla ^\alpha _0)^2\prod \limits _{n=1}^{\infty}
\left[\det (ig^{-1}\nabla ^H_n)^2\prod \limits _{\alpha > 0}
 \det (-ig^{-1}\nabla ^\alpha _n)^2\right]=\nonumber \\
&=& (2\pi l)^r\prod \limits _{\alpha > 0}(a,\alpha )^4\prod
\limits _{n=1}^{\infty} \left[ \left( \frac{n}{gl}\right)^{4r}
\prod \limits _{\alpha > 0}\left(
\frac{n^2}{g^2l^2}-(a,\alpha )^2\right)^4\right].
\ea
Set $(\det g_{AB})^{1/2}=C(l)\mu (a),\ \ \mu (a)=\kappa ^2(a)$.
Including all divergences of the product (6.11) into $C(l)$ we get \cite{pisa}
\ba
\kappa (a)&=& \prod \limits _{\alpha > 0}\left[
\frac{\pi (a,\alpha)}{a_0}
\prod \limits _{n=1}^{\infty}\left(1-\frac{(a,\alpha
)^2}{a^2_0n^2}\right)\right]=
\prod \limits _{\alpha > 0}\sin \frac{\pi (a,\alpha )}{a_0},\\
C(l)&=& (2\pi l)^{r/2}\left (\frac{a_0}{\pi }\right ) ^{N_+}\prod
\limits _{n=1}^{\infty}(n^2a_0^2)^{r+2}\ ,
\ea
where $a_0=(gl)^{-1},\ \ N_+=(N-r)/2$ is the number of positive roots in $X$;
 the last equality in (6.12) results from a product formula given in p.37,
\cite{ryz}. Following the prescription (B.13) we define the scalar product
in the model
\be
\int\limits_{[A]}\prod\limits_{x\in {\bf s}^1}dA(x)\Phi^*_n[A]
\Phi_{n'}[A]\rightarrow \int\limits_{K_W^+}da\kappa^2(a)
\Phi_n^*(a)\Phi_{n'}(a)=\delta_{nn'}\ ,
\ee
where the infinite constant $C(l)\int_{G/G_H}\prod_x\wedge dw(x)$
is removed by a renormalization of physical states, which we denote
by the arrow in (6.14). This.  renormalization is admissible because of
the gauge invariance of physical states $\Phi_n[A]=\Phi_n(a)$ (compare
(3.3) and (6.3)). Notice that Eq.(6.2) being written via the new variables
has the form
\be
\hat{\sigma }\Phi _n[a,\omega ]=
-ig\hat{\Omega}\frac{\delta }{\delta w}\Phi_n[a,w] =0\ ;
\ee
here we have used the notation $\hat{\Omega}y=\Omega y\Omega^{-1}=
y\hat{\Omega}^T,\ \hat{\Omega}^T\hat{\Omega}= \hat{\Omega}\hat{\Omega}^T
= 1$ for any element $y\in X$. So, $\det\hat{\Omega}\neq 0$ and solutions
to (6.15) are given by functionals independent of $w(x)$.

To prove the equivalence of Eqs.(6.2) and (6.15), we derive first the
following relations from (6.5)
\ba
da&=&P_0^H\hat{\Omega}^T\delta A\ ,\\
\delta w&=&ig\nabla^{-1}(a)(1-P_0^H)\hat{\Omega}^T\delta A
\ea
with $P_0^H$ being a projector on ${\cal F}_0^H$ (the operator
$\nabla(a)$ is invertible on $(1-P_0^H){\cal F}$). The straightforward
calculations lead us to the desired result
\ba
\nabla(A)\frac{\delta}{\delta A}&=&\nabla(A)\left[
\left(\frac{\delta a}{\delta A},\frac{\pl}{\pl a}\right)_a +
\langle\frac{\delta w}{\delta A},\frac{\delta }{\delta w}\rangle_w
\right] =\\
&=&\nabla(A)\left[\left(P_0^H\hat{\Omega}^T\right)^T\frac{\pl}{\pl a}
+\left(ig\nabla^{-1}(a)(1-P_0^H)\hat{\Omega}^T\right)^T
\frac{\delta}{\delta w}\right]=\\
&=&\hat{\Omega}\nabla(a)P_0^H\frac{\pl}{\pl a}
-ig\hat{\Omega}\nabla(a)(1-P_0^H)\nabla^{-1}(a)\frac{\delta}{\delta w}
=-ig\hat{\Omega}^T\frac{\delta}{\delta w}\ .
\ea
In (6.18), the suffix at the scalar product brackets denotes
variables over whose indices the scalar product is taken, i.e.
all indices of $A(x)$ (the Lie algebra ones and $x\in {\bf S}^1$)
 in the scalar products entering into (6.18)
are left free. Equality (6.19) results from (6.16-17). In (6.20),
we have used $\nabla(a)P_0^H\pl/\pl a\equiv 0$ and $\nabla(A)
\hat{\Omega}=\hat{\Omega}\nabla(a)$.

Thus, we have proved the gauge
invariance of the variable $a$, $[\hat{\sigma },\hat{a}]=0$. In this
approach, the Gauss law (6.2) is explicitly solved (even in the 4D case
\cite{gr}). We do not fix a gauge at all, but we do choose a certain
coordinates on the gauge orbit space.

To project the functional Laplace operator in (6.1) on the gauge
orbit space spanned by variables $a$, one should calculate the Laplace-Beltrami
operator in the new coordinates (6.3) and drop all terms containing
$\delta/\delta w$ in it. A general recipe is given by (B.11).
Using it, we arrive at a quantum mechanical problem
\be
\hat{H}^{ph}\Phi_n(a)=\left[-\frac{\hbar^2}{4\pi l}\frac{1}{\kappa(a)}
(\pl_a,\pl_a)\circ\kappa(a) -E_C\right]\Phi_n(a)=E_n\Phi_n(a)\ ,
\ee
where we have taken into account $g^{11}= (2\pi l)^
{-1}$, the
quantum potential (B.12) turns out to be independent of $a$,
\be
V_q=\frac{\hbar^2}{4\pi l}\kappa^{-1}
(\pl_a,\pl_a)\kappa =-
\frac{\pi\hbar^2}{4a_0^2l}\left(\sum\limits_{\alpha>0}\right)^2 =-E_C\ .
\ee
A proof of (6.22) is given in Appendix C. In fact, $E_C$ coincides with
the Casimir energy related to the curvature of the group manifold
(cf. the SU(N) case considered in \cite{wad},\cite{het}). Notice that
the Casimir energy naturally appears upon solving the constraint (6.2),
which would not occur through quantizing after eliminating all unphysical
degrees of freedom by imposing the gauge condition $A_1(x)=a$ (this
approach is considered in \cite{sem} for $G=SU(N)$).
In the case of an arbitrary gauge, a quantization of the
gauge-fixed theory might give rise to not only loosing the Casimir
energy, but even to more drastic consequences -- the gauge dependence
of the quantum theory -- as we have shown in Appendix B. To ensure
the gauge independence, one should quantize before removing gauge
degrees of freedom.

As has been proved in Appendix B, any regular solution of (6.21)
must be invariant with respect to the discrete transformations (3.17)
$S_F=W_A$
\be
\Phi_n(\hat{s} a) = \Phi_n(a)\ ,\ \ \ \ \ \ \ \hat{s}\in W_A\ .
\ee
In Appendix D we verify (6.23) by explicit solving Eq.(6.21) and show that
the eigen functions are given by characters of the irreducible representations
of $G$, while the eigen values are proportional to eigen values of the
quadratic Casimir operator of $G$. The property (6.23) determines the
boundary conditions which we need to construct the PI representation of the
transition amplitude by means of the method of "reflected"
trajectories. It gives also an analytical continuation of physical wave
functions to
the unphysical region $a\in H$ (to the covering space of the gauge orbit space)
in full accordance with a general analysis in Appendix B (see (B.14)).

To obtain a PI representation of the transition amplitude
$U^{ph}_t(a,a')= \langle a|e^{-it\hat{H}^{ph}/\hbar}|a'\rangle$,
one should repeat calculations (B.17-30) for this particular model.
Due to (6.23) and (6.14), the analytical continuation of the unit
operator kernel reads
\be
\la a|a'\ra = \sum\limits_{\hat{s}\in W_A}\left(\kappa(a)\kappa(\hat{s}a')
\right)^{-1}\delta^r(a-\hat{s}a')\ ,\ \ \ \ \ a\in H,\ \ a'\in K_W^+\ .
\ee
The kernel (6.24) is $W_A$-invariant. Consider the reflection (3.17); we have
to prove $\la \hat{s}_{\alpha ,n}a|a'\ra = \la a|a'\ra$. Change the summation
over $\hat{s}$ in (6.24) by $\hat{s}_\alpha \hat{s}$ and use relations (C.7-8)
$\kappa(\hat{s}_{\alpha ,n}a) =\kappa (\hat{s}_\alpha a)=-\kappa (a)$
and $\delta ^r(\hat{s}_{\alpha ,n}a-\hat{s}_\alpha \hat{s}a')=\delta (\hat{s}
_\alpha ^{-1}\hat{s}_{\alpha ,n}a-\hat{s}a')$. The element $\hat{s}_
\alpha ^{-1}\hat{s}_{\alpha ,n}$ is a translation through periods of the unit
group lattice which are independent of $a$. Since the measure $\kappa (a)$
is invariant under these translations (see Appendix C), the shift of
$\hat{s}^{-1}_\alpha
\hat{s}_{\alpha ,n}a=a-2na\alpha/(\alpha ,\alpha ),\ n\in \Z$, can be
included into shift of $a'$ generated by all $\hat{s}\in W$ by virtue of
resummation in the infinite series (6.24), which completes
the proof.

Therefore the infinitesimal transition amplitude is given by (B.18) where
$(\mu \mu '')^{-1/2}$ is changed by $(\kappa \kappa '')^{-1},\ \R^M
\rightarrow H\sim \R^r\in a''$,
\be
Q(a'',a')=\sum_{\hat{s}\in W_A}\delta^r(a''-\hat{s}a')\ ,
\ee
and in (B.19) $(u,p)\rightarrow (a,p_a),\ \ p_a\in H\sim \R ^r$,
\be
H^{eff}(a,p_a)=\frac{1}{4\pi l}(p_a,p_a)-E_c\ ;
\ee
notice that the second term in (B.20) vanishes because $g_{ph}$ is
independent of $a$. The $W_A\ -$ invariance of the infinitesimal kernel
\be
U^{ph}_\epsilon (\hat{s}_{\alpha,n},a')=U^{ph}_\epsilon (a,a'),\ \
\hat{s}_{\alpha,n}\in W_A\ ,
\ee
results from the representation (B.18) for $U^{ph}_\epsilon
(\hat{s}_{\alpha ,n}a,a')$. Indeed, the transformation $\hat{s}_{\alpha ,n}$ is
a
composition of a translation $t_n\in T_e$ and a reflection
$\hat{s}_\alpha$ from the Weyl group. The $T_e$ invariance of
$U^{ph}_\epsilon (a,a')$ is obvious because $U^{eff}_\epsilon (a,a')$ in
(B.18) depends only on $\Delta =a-a''$. Indeed, any $T_e$-shift of $a$ can be
removed by a shift of the integration variable $a''$. The $T_e -$invariance
follows from the $T_e-$invariance  of the kernel (6.25) and the function
$\kappa (a)$ (see appendix C). To prove the $W\ -$invariance, one
should do the integral over $a''$ in (B.18), then change the integration
variables in (B.19) $p_a\rightarrow \hat{s}_\alpha p_a$ and use the
$W$-invariance of the Killing form
$(\hat{s}_\alpha p_a,\hat{s}_\alpha a)=(p_a,a)$.

Proved (6.27) we find for the convolution of infinitesimal evolution
operator kernels
\ba
U^{ph}_{2\epsilon }(a,a')&=&\int\limits _{K^+_W}da_1\kappa ^2(a_1)U^{ph}
_\epsilon (a,a_1)U^{ph}_\epsilon (a_1,a')=\\
&=& \sum _{W_A}\int\limits _{K^+_W}\frac{da_1\kappa ^2(a_1)}{\kappa (a)\kappa
(\hat{s}a_1)} U^{eff}_\epsilon (a,\hat{s}a_1)U^{ph}_\epsilon
(a_1,a')=\\
&=& \int\limits _{H\sim \R^r}\!\!\frac{da''}{\kappa (a)\kappa (a'')}\left (
\ \int\limits _{H\sim \R^r} da_1U^{eff}_\epsilon(a,a_1)U^{eff}_\epsilon
(a_1,a'')\right ) Q(a'',a')\ ;
\ea
where (6.29) is obtained by substitution the representation (B.18) into
(6.28) and doing the integral over $a''$, then we have changed the integration
variables in (6.29) $a_1\rightarrow \hat{s}a_1$ (the Jacobian being
$J_s(a)=\det \hat{s}=\pm 1$), and used (6.27), the integration rule
$\sum _{W_A}^{}\int _{\hat{s}K_W^+}^{}da_1=\int _{H}^{}da_1
,\ \ \kappa^2(\hat{s}a)=\kappa^2(a)$ and (B.18)
again to derive (6.30). Equalities (6.28-30) are to illustrate (B.26-28).
Thus, for a finite time the convolutions (B.29) and (B.30) yield
\ba
U^{ph}_t(a,a')&=& \int\limits_{\R^r}^{}\frac{da''}{\kappa (a)\kappa
(a'')}U^{eff}_t(a,a'')Q(a'',a)\\
U^{eff}_t(a,a'')&=& \int\limits_{\R^{2r}}^{}\prod\limits_{\tau =0}^{t}
\left( \frac{dp_a(\tau ) da(\tau)}{(2\pi\hbar )^r}\right)\exp
\frac{i}{\hbar}\int\limits_{0}^{t}d\tau\left((p,\dot{a})-\frac{(p,p)}
{4\pi l}+E_C\right)=\\
&=&\left(\frac{l}{i\hbar t}\right)^{r/2}\exp \left(\frac{i\pi
l(a-a'')^2}{\hbar t}+\frac{it}{\hbar}E_C\right)\ .
\ea

The function (6.33) is a transition amplitude for a free $r$-dimentional
particle of mass $2\pi l$. In fact, we have proved that the $2D$
Yang-Mills theory on a cylindrical spacetime is equivalent to a quantum
theory of an $r$-dimentional free particle moving in a polyhedron being
the Weyl cell of the gauge group. Substituting $\Phi _n=\kappa
^{-1}\Psi _n$ into (6.21) we see that $\Psi _n$ satisfies the Schroedinger
equation for a free $r$-dimentional particle. But it is not completely
free motion because of the boundary condition (6.23). The particle
"interacts" with the boundary of the Weyl cell, which leads to the
"twisted" boundary condition
\begin{equation}
\Psi _n(\hat{s}a)=\det \hat{s} \Psi _n(a),\ \ \hat{s}\in W_A\ .
\end{equation}
Any regular function on $H$ obeying (6.34) vanishes on $\pl K^+_W$ (even
on the whole
diagram $D(X)$ (see Sec.3)) providing a regular behavior of $\Phi
_n$ on $\pl K^+_W$ (and on the covering space $H$).
An important point following from our consideration is
that the boundary condition (6.23) or (6.34) automatically results from the
Dirac method, and we do not need to impose them (or another ones) by hand
(in contrast with \cite{het}, \cite{sem}). Moreover, $W_A-$invariant
functions (6.23) are regular functions of the Wilson loop (see Appendix D),
i.e. they turn out to be explicitly gauge-invariant.

Notice also that (6.31) can be regarded as a transition amplitude
between two Wilson loops because a space of Wilson loops is isomorphic
to the Weyl cell $K_W^+$ (cf. (D.6)).

\vskip 0.8cm
\noindent
{\Large\bf Acknowledgments}

\vskip 0.8cm
The author is grateful to J.-B.Zuber and D.Bernard for useful discussions.

\vskip  0.8cm
\renewcommand{\theequation}{A.\arabic{equation}}
\noindent
{\Large\bf Appendix}

\vskip 0.8cm
\noindent
{\large\bf A.\  The Cartan-Weyl basis in Lie algebras}
\setcounter{equation}0

\vskip 0.8cm
Any simple Lie algebra $X$ is characterized by a set of linearly
independent $r$-di\-men\-sio\-nal vectors $\vec{\omega}_j,\ j= 1,2,...
, r= rank\ X$, called simple roots. The simple roots form a basis
for the algebra root system. Any root $\vec{\alpha}$ is a linear
combination of $\vec{\omega}_j$ with either non-negative integer
coefficients ($\vec{\alpha}$ a positive root) or non-positive
integer coefficients ($\vec{\alpha}$ a negative root). Obviously,
all simple roots are positive. If $\vec{\alpha}$ is a root then
$-\vec{\alpha}$ is also a root. The root system is fixed by the
Cartan matrix $c_{ij}=-2(\vec{\omega}_i,\vec{\omega}_j)/
(\vec{\omega}_j,\vec{\omega}_j)$ (here $(\vec{\omega}_i,
\vec{\omega}_j)$ is a usual scalar product of two $r$-vectors)
 which can be uniquely restored from the Dynkin diagrams \cite{hel}.
Elements of the Cartan matrix are integers.
For any two roots $\vec{\alpha}$ and $\vec{\beta }$, the cosine of the
angle between them can take only following values $(\vec{\alpha},
\vec{\beta})[(\vec{\alpha},\vec{\alpha}) (\vec{\beta},\vec{\beta}
)]^{-1/2} = 0,\pm 1/2,\pm 1/\sqrt{2},
\pm \sqrt{3}/2$. By means of this fact and
the Cartan matrix, the whole root system can be restored
\cite{hel}, p.460.

For any two elements $x,\ y$ of $X$, the Killing form is defined as
$(x,y) = tr(\hat{x}\hat{y}) = (y,x)$ where the operator $\hat{x}$
acts in $X$, $\hat{x}y =[x,y]$. A maximal Abelian subalgebra $H$
in $X$ is called the Cartan subalgebra, $\dim H= rank\ X =r$. There
are $r$ linearly independent elements $\omega_j$ in $H$ such that
$(\omega_i,\omega_j) = (\vec{\omega}_i,\vec{\omega}_j)$. We shall
also call the algebra elements $\omega_i$ simple roots. It will
not lead to any confusing in what follows because the root space $\R^r$ and the
Cartan
subalgebra are isomorphic, but we shall keep arrows over elements of
$\R^r$. The corresponding elements of $H$ have no the over-arrow.

A Lie algebra $X$ is decomposed into the direct sum $X=
H\oplus\sum_{\alpha >0} (X_\alpha\oplus X_{-\alpha}),\ \alpha$ ranges
the positive roots, $\dim X_{\pm\alpha} = 1$. Simple roots form a
basis (non-orthogonal) in $H$. Basis elements $e_{\pm\alpha}$ of
$X_{\pm\alpha}$ can be chosen such that \cite{hel}, p.176,
\begin{eqnarray}
\left[e_\alpha ,e_{-\alpha}\right]& =& \alpha\ ,\\
\left[h,e_\alpha \right] &=& (\alpha, h)e_\alpha\ ,\\
\left[e_\alpha,e_\beta\right] &=& N_{\alpha,\beta}e_{\alpha+\beta}\ ,
\ea
for all $\alpha, \beta$ belonging to the root system and for any $h
\in H$, where the constants $N_{\alpha,\beta}$ satisfy
$N_{\alpha,\beta}= -N_{-\alpha,-\beta}$. For any such choice
$N_{\alpha,\beta}^2 = 1/2q(1-p)(\alpha,\alpha)$ where $\beta +n\alpha\
(p\leq n \leq q)$ is the $\alpha$-series of roots containing $\beta$;
$N_{\alpha,\beta}=0$ if $\alpha+\beta$ is not a root.

Any element $x\in X$ can be decomposed over the Cartan-Weyl basis
(A.1-3),
\be
x = x_H + \sum_{\alpha >0}(x^\alpha e_\alpha + x^{-\alpha}e_{-\alpha})
\ee
with $x_H$ being the Cartan subalgebra component of $x$.

The commutation relations (A.1-3) imply a definite choice of the norms
of the elements $e_{\pm\alpha}$, namely, $(e_{\pm\alpha},
e_{\pm\alpha})= 0$ and $(e_\alpha,e_{-\alpha})= 1$ \cite{hel}, p.167.
Norms of simple roots are also fixed in (A.1-3). Consider, for
instance, the $su(2)$ algebra. There is just one positive root
$\omega$. Let its norm be $\gamma = (\omega,\omega)$. The Cartan-Weyl
basis reads $[e_\omega,e_{-\omega}] = \omega$ and
$[\omega,e_{\pm\omega}] = \pm \gamma e_{\pm\omega}$. Let us calculate
$\gamma$ in this basis.  By definition $\gamma =
tr\hat{\omega}\hat{\omega}$. The operator $\hat{\omega}$ is $3\times
3$ diagonal matrix with $0,\pm\gamma$ being its diagonal elements as
follows from the basis commutation relations and the definition of the
operator $\hat{\omega}$. Thus, $tr\hat{\omega}^2 = 2\gamma^2 =\gamma$,
i.e. $\gamma =1/2$.

The $su(3)$ algebra has two equal-norm simple roots $\vec{\omega}_1$
and $\vec{\omega}_2$ with the angle between them equal to $2\pi/3$.
For the corresponding Cartan subalgebra elements we have $(\omega_1,
\omega_1)= (\omega_2,\omega_2) =\gamma$ and $(\omega_1,\omega_2) =
-\gamma/2$. The whole root system is given by six elements
$\pm\omega_1,
\pm\omega_2$ and $\pm(\omega_1+\omega_2) \equiv \pm \omega_{12}$. It is
readily to see $(\omega_{12},\omega_{12})= \gamma$ and $(\omega_1,
\omega_{12})= (\omega_2,\omega_{12}) = \gamma/2$. All the roots have
the same norm and the angle between two neighbor roots is equal to
$\pi/3$. Having obtained the root pattern, we can evaluate the number
$\gamma$. The basis (non-orthogonal) consists of eight elements
$\omega_{1,2}, \ e_{\pm 1},\ e_{\pm 2}$ and $ e_{\pm 12}$ where we
have introduced simplified notations $e_{\pm\omega_1} \equiv e_{\pm
1}$, etc. The operators $\hat{\omega}_{1,2}$ are $8\times 8$ diagonal
matrices as follows from (A.2) and $[\omega_1,\omega_2]= 0$. Using
(A.2) we find $tr\hat{\omega}_{1,2}^2 = 3\gamma^2 = \gamma$ and,
therefore, $\gamma= 1/3$. As soon as root norms are established, one
can obtain the structure constants $N_{\alpha,\beta}$. For $X=su(3)$
we have $N_{1,2}^2 = N_{12,1}^2 =N_{12,2}^2 = 1/6$ and all others
vanish (notice that $N_{\alpha,\beta} =-N_{-\alpha,-\beta}$ and
$N_{\alpha,\beta}= -N_{\beta,\alpha}$). The latter determines the
structure constants up to a sign. The transformation
$e_\alpha\rightarrow -e_\alpha,\ N_{\alpha,\beta} \rightarrow
-N_{\alpha,\beta}$ leaves the Cartan-Weyl basis untouched. Therefore
only relative sings of the structure constants must be fixed.
Fulfilling the Jacobi identity for elements $e_{-1},\ e_1,\ e_2$ and
$e_{-2},\ e_1,\ e_2$ results in $N_{1,2}= -N_{12,-1}$ and $N_{1,2} =
N_{12,-2}$, respectively. Now one can set $N_{1,2}=N_{12, -2} =
-N_{12,-1} = 1/\sqrt{6}$, which completes determining the structure
constants of $su(3)$.

One can construct a basis orthonormal with respect to the Killing
form. With this purpose we introduce the elements \cite{hel}, p.181,
\be
s_\alpha = i(e_\alpha - e_{-\alpha})/\sqrt{2},\ \ \ \ \ \ c_\alpha =
(e_\alpha + e_{-\alpha})/\sqrt{2}\ .
\ee
Then $(s_\alpha,s_\beta) = (c_\alpha,c_\beta) = \delta_{\alpha\beta}$
and $(c_\alpha,s_\beta) = 0$. Also,
\be
(x,x)=\sum_{\alpha >0} \left[(x^\alpha_s)^2 + (x_c^\alpha)^2\right] +
(x_H,x_H)\ ,
\ee
where $x_{s,c}^\alpha$ are real decomposition coefficients of $x$ in the
orthonormal basis (A.5). Supplementing (A.5) by an orthonormal basis
$h_j,\ (h_j,h_i)=\delta_{ij}$, of the Cartan subalgebra (it might be
obtained by orthogonalizing the simple-root basis of H), we get an
orthonormal basis in $X$; we shall denote it $\lambda_a = (h_j,
s_\alpha, c_\alpha)$.

Suppose we have a matrix representation of $X$. Then $(x,y)= c_r\ tr\
xy$ where $xy$ means a matrix multiplication. The number $c_r$ depends
on $X$.  For classical Lie algebras, the numbers $c_r$ are listed in
\cite{hel}, pp.187-190. For example, $c_r = 2(r+1)$ for $X= su(r+1)$.
Using this, one can establish a relation of the orthonormal basis
constructed above for $su(2)$ and $su(3)$ with the Pauli matrices
\cite{hua}, p.13, and the Gell-Mann matrices \cite{hua}, p.17,
respectively. For the Pauli matrices we have
$[\tau_a,\tau_b]=2i\varepsilon_{abc}\tau_c,\
\varepsilon_{abc}$ a totally antisymmetrical tensor, $\varepsilon_{123}=1$,
hence, $(\tau_a,\tau_b) = -4\varepsilon_{ab'c'}\varepsilon_{bc'b'} = 8
\delta_{ab} = 4 tr\ \tau_a\tau_b$ in full accordance with $c_r= 2(r+1),\
r=1$. One can set $\omega = \tau_3/4,\ s_\omega= \varphi\tau_1$ and
$c_\omega = \varphi\tau_2$ where $1/\varphi = 2\sqrt{2}$. A similar
analysis of the structure constants for the Gell-Mann matrices
$\lambda_a$
\cite{hua}, p.18, yields $\omega_1 = \lambda_3/6,\ s_1 =\varphi\lambda_1,\
c_1= \varphi\lambda_2,\ \omega_2= (\sqrt{3}\lambda_8 - \lambda_3)/12,\
s_2= \varphi\lambda_6,\ c_2= \varphi\lambda_7\ , \omega_{12} =
(\sqrt{3}\lambda_8 + \lambda_3)/12,\ s_{12}=\varphi\lambda_5$ and
$c_{12}= -\varphi\lambda_{4}$ where $1/\varphi = 2\sqrt{3}$. This
choice is not unique. Actually, the identification of non-diagonal
generators $\lambda_a,\ a\neq 3,8$ with (A.5) depends on a
representation of the simple roots $\omega_{1,2}$ by the diagonal
matrices $\lambda_{3,8}$.  One could choose $\omega_1=\lambda_3/6$ and
$\omega_2=-(\sqrt{3}\lambda_8 + \lambda_3)/12$, which would lead to
another matrix realization of elements (A.5).

\renewcommand{\theequation}{B.\arabic{equation}}
\vskip 0.8cm
\noindent
{\large\bf B. \ The Hamiltonian path integral in an arbitrary gauge}

\setcounter{equation}0
\vskip 0.8cm

Consider a quantum theory determined by the Schroedinger equation
\be
\left(-\frac12\langle\frac{\pl}{\pl x},\frac{\pl}{\pl x}\rangle
+ V(x)\right)
\psi_E = E\psi_E\ .
\ee
The eigen-functions $\psi_E$ are normalized by the condition
\be
\int\limits_{\R^N}dx\psi_E^*(x)\psi_E(x)= \delta_{EE'}\ .
\ee
We assume $x$ to realize a linear representation of a compact group $G$:
$x\rightarrow \Omega(\omega)x,\ \Omega(\omega)\in G$ and $V(\Omega x)=V(x)$;
$\langle x,y\rangle = \sum_1^Nx_iy_i = \langle \Omega x,\Omega y\rangle$ is an
invariant
scalar product in the representation space that is isomorphic to $\R^N$. The
theory turns into the gauge one if we require that physical states are
annihilated by operators $\hat{\sigma}_a$
generating $G$-transformations of $x$,
$\hat{\sigma}_a\Phi(x)=0$. These conditions determine a physical
subspace in the Hilbert space. By definition, we have $\exp(\omega_a
\hat{\sigma}_a)\psi(x) = \psi(\Omega(\omega)x) $ where $\Omega(\omega) \in G$.
Therefore, the physical states are $G$-invariant
\be
\Phi(\Omega(\omega)x)= \Phi(x)\ .
\ee

Let a number of physical degrees of freedom in the system is
equal to $M$, then a number of independent constraints is $N-M$.
Suppose we would like to span the physical configuration space
$K\sim \R^N/G$ by coordinates ranging a gauge condition surface
$F(x)=0$. We assume the gauge condition to be complete, meaning that
there is no unphysical degree of freedom left. Let $u\in \R^M$ be
a parameter of the gauge condition surface; $x=f(u)$ such that
$F(f(u))$ identically vanishes for all $u\in \R^M$. By analogy
with (4.15) we introduce curvilinear coordinates
\be
x=x(\theta,u)=\Omega(\theta)f(u)\ ,
\ee
where variables $\theta$ ran over the manifold $G/G_f$ with $G_f$
being a stationary group of the vector $x=f,\ G_ff=f$.

The metric tensor in the new coordinates reads
\be
\langle dx,dx \rangle = \langle df, df \rangle + 2\langle df,d\theta f
\rangle + \langle d\theta f, d\theta f\rangle \equiv g_{AB}dy^Ady^B\ ,
\ee
where we have put $d\theta =\Omega^ +d\Omega $ and $dy^1\equiv du,\ dy^2\equiv
d\theta$. Therefore,
\be
\int\limits_{\R^N}dx = \int\limits_{G/G_f}\wedge d\theta
\int\limits_K d^M u\mu(u)\ ;
\ee
here $\mu(u)= (\det g_{AB})^{1/2},\ K$ is a subdomain in $\R^M$ such
that the mapping (B.4), $K\otimes G/G_f\rightarrow \R^N$, is one-to-one.
To determine $K$, one should find transformations $\theta,u
\rightarrow \hat{s}\theta, \hat{s}u,\ \hat{s}\in \tilde{S}_F$ which
leave $x$ untouched, $x(\hat{s}\theta, \hat{s}u)=x(\theta,u)$ (cf.
(4.19)). Obviously, $\tilde{S}_F=T_e\times S_F$ where $T_e$ is a
group of translations of $\theta$ through periods of the manifold
$G/G_f$, while the set $S_F$ is obtained by solving Eqs.(4.2-3) with
${\bf f}\rightarrow f\in \R^N,\ u\in R^M,\ \Omega_s\in G$, so
$K\sim \R^M/S_F$. Indeed, assuming
Eq.(4.2) to have non-trivial solutions (the trivial one $\Omega _s=1$ always
exists by the definition of $f(u)$) we observe that all points
$\Omega _sf$ belong to the gauge condition surface and, hence, $\Omega _sf(u)=
f(u_s), \ u_s=u_s(u)$. Consider transformations of $\theta$ generated
by the group shift $\Omega (\theta)\rightarrow \Omega (\theta)T^{-1}_s
=\Omega (\theta_s),\
\theta_s=\theta_s(\theta,u)$. Setting $\hat{s}u=u_s$ and $\hat{s}
\theta= \theta_s$ we see that the transformations $\hat{s}\in S_F$
leave $x=x(\theta,u)$ untouched. To avoid a "double" counting in the
scalar product integral (B.6),
one has to restrict the integration domain for $u$ to the
quotient $\R^M/S_F=K$. The modular domain $K$ can also be determined
by the requirement as a part of the gauge condition surface $x=f(u),\
u\in K\subset \R^M$, which has just one common point with any gauge orbit.

A choice of the fundamental domain parametrization
is not unique as we have seen in Sec.4 (cf. (4.5-7) and (4.9-11)).
In (B.6), $\mu >0$ for $u\in K$. Having chosen the parametrization
of $K$, we fix a representation of $S_F$ by functions
$\hat{s}u=u_s(u),\ u\in K,\ u_s\in K_s,\ K_s\cap K_{s'}=\emptyset$ for
any $\hat{s}\neq\hat{s}'$ and $\R^M= \cup_s K_s$ up to a set of zero
measure being a unification of the boundaries $\pl K_s$. We define
an orientation of $K_s$ so that for all $\hat{s}\in S_F$,
$\int_{K_s}du\phi\geq 0$ if $\phi\geq 0$, so that the following
rules hold
\ba
\int\limits_{\R^M}du &=& \sum\limits_{S_F}\int\limits_{K_s}du\ ,\\
\int\limits_Kdu |J_s(u)|&=& \int\limits_{K_s}du\ ,
\ea
where $J_s(u) = Du_s/Du$ is the Jacobian, the absolute value of
$J_s$ has been inserted into the right-hand side of (B.8) for
preserving the positive orientation of the integration domain.

{\em Remark}. A number of elements in $S_F$ can depend on $u$.
We define a region $\R^M_\alpha\subseteq \R^M $ such that
$S_F=S_\alpha$ has a fixed number of elements for all $u\in
\R^M_\alpha$. Then $K=\cup_\alpha K_\alpha,\ K_\alpha
=\R^M_\alpha/S_\alpha,\ \R^M= \cup_\alpha\R^M_\alpha$. The
sum in (B.7) implies $\sum_{S_F}=\sum_\alpha\sum_{S_\alpha}$
and $K_s$ in (B.7-8) carries an additional suffix
$\alpha$. In what follows we shall omit it and use the simplified
notations (B.7-8) to avoid complications of formulas. The suffix
$\alpha$ can be easily restored  by means of the rule
proposed above.

For the mechanical model considered in Sec.4, we have $G=SO(2),
G_f=1,\ \det G_{AB} = {\bf f}'^2{\bf f}^2 -({\bf f}'T{\bf f})^2
= ({\bf f}',{\bf f})^2 = \mu^2(u)$. Set $K=\cup_\alpha K_\alpha,\
K_1=(0,u_o/\gamma_0),\ K_2=(u_0/\gamma_0, u_0),\ K_3$\
$ = (u_0,\infty)$,
i.e. $K=\R_+$, then $\int_{-\infty}^\infty du =
\sum_\alpha\int_{\R_\alpha}du$ and (B.7) means that the upper
integral limit is always greater than the lower one, for example,
$$
\int\limits_{\R_2}du =\left(\int\limits_{-3u_0}^{-2u_0}+
\int\limits_{-2u_0}^{-u_0} +
\int\limits_{-u_0}^{-u_0/\gamma_0}+
\int\limits_{u_0/\gamma_0}^{u_0}\right) du\ ,
$$
where the terms of the sum correspond to integrations over
$\hat{s}_3K_2,\ \hat{s}_2K_2,\ \hat{s}_1K_2$ and $K_2$,
respectively (cf. (4.5-7)). The following chain of equalities
is to illustrate the rule (B.8)
\be
\int\limits_{\hat{s}_3K_2} du_{s_3} = \int\limits_{-3u_0}^{-2u_0}
du_{s_3} = \int\limits_{u_0}^{u_0/\gamma_0} du J_{s_3} =
- \int\limits_{u_0/\gamma_0}^{u_0}du J_{s_3}= \int\limits_{K_2}
du |J_{s_3}|\ ;
\ee
the last equality results from $J_{s_3}= du_{s_3}/du < 0$
(cf. (4.7)).

By means of the curvilinear coordinates (B.4) we can naturally
 incorporate a gauge condition chosen into the Dirac operator
method \cite{dir} of quantizing first-class constrained systems.
Solutions of the equation $\hat{\sigma}_a\tilde{\Phi}(x)=0$ are
given by functions independent of $\theta$,
\be
\tilde{\Phi}(x)=\tilde{\Phi}(\Omega (\theta)f(u))=
\tilde{\Phi}(f(u))=\Phi(u)\ ,
\ee
because $\hat{\sigma}_a$ generate shifts of $\theta$ and leave
$u$ untouched. To obtain a physical Hamiltonian, one has to write
the Laplacian in (B.1) via the new variables (B.4) and omit all
terms containing derivatives with respect to $\theta$. In so doing,
we get \cite{jpa},\cite{book}
\be
\hat{H}^f_{ph}\Phi_E(u)=
\left(\frac12\hat{p}_ig^{ij}_{ph}\hat{p}_j + V_q(u) +
V(f(u))\right)\Phi_E(u) = E\Phi(u)\ ;
\ee
here we have introduced hermitian momenta $\hat{p}_i = -i\mu^{-1/2}
\pl_j\circ \mu^{1/2},\ \pl_j =\pl/\pl u^j$;
the metric $g^{ij}_{ph}$ in the physical
configuration space is the $11$-component of a tensor $g^{AB}$ inverse
to $g_{AB},\ g^{AC}g_{CB}=\delta^A_B,\ g^{ij}_{ph}= (g^{11})^{ij},\
i,j=1,2,...,M$; a quantum potential
\be
V_q=\frac{1}{2\sqrt{\mu}}(\pl_ig^{ij}_{ph})\pl_j\sqrt{\mu} +
\frac{1}{2\mu}g^{ij}\pl_i\pl_j\sqrt{\mu}\
\ee
appears due to the chosen ordering of
the operators $\hat{u}^i$ and $\hat{p}_i$ in the Laplace-Beltrami
operator. The scalar product is reduced to
\be
\int\limits_{\R^N}dx\Phi^*_E(u)\Phi_{E'}(u)\rightarrow
\int\limits_Kd^Mu \mu(u)\Phi^*_E(u)\Phi_{E'}(u)=\delta_{EE'}\ ,
\ee
where a gauge orbit volume (integral over $G/G_f$ (see (B.6)) has
been included into norms of physical states, which we denoted by
the arrow in (B.13). A construction of an operator
description of a gauge theory in a given gauge condition is completed.

Notice, in this approach the variables $u$ appear to be gauge-invariant;
they parametrize the physical configuration space $CS_{ph}=\R^N/G$.
Two different choices of $f(u)$ correspond two different parametrizations
of $CS_{ph}$ related to each other by a change of variables
$u=u(\tilde{u})$ in (B.11-13). Therefore quantum theories with different
$f$'s are unitary equivalent \cite{book}.

To illustrate this statement, we consider again the simplest case
$G=SO(2),\ M=1,\ g_{ph}=r^2(u)/\mu^2(u)$, and compare descriptions
in the coordinates (4.15) and in the polar ones ($f_1=r, f_2=0$).
With this purpose we change variables $r=r(u)$ in (B.11-13). For
$u\in K$ the function $r(u)$ is invertible, $u=u(r), r\in \R_+$.
Simple straightforward calculations \cite{gr} lead us to the
following equalities $\hat{H}^f_{ph}=1/2\hat{p}_r^2 + V_q(r)
+ V,\ \hat{p}_r=-i r^{-1/2}\pl_r \circ r^{1/2},\ V_q
=-(8r^2)^{-1},\ \int_K du\mu = \int_0^\infty drr$. It is nothing
but quantum mechanics of a radial motion on a plane. All theories
with different $f$'s are unitary equivalent to it and, therefore,
to each other. One should stress that the operator ordering we
obtained by applying the Dirac method plays the crucial role
in providing this unitary equivalence. Another ordering of
operators in (B.11) would break this property.

A few observations resulting from our consideration have to
be emphasized.

1. All regular solutions of (B.11) have a unique analytical
continuation  to the whole space $u\in \R^M$, and they are
$S_F$-invariant,
\be
\Phi_E(u_s(u))=\Phi_E(u)\ ,\ \ \ \ \ \  \ u\in K\ .
\ee
For a proof, we point out that any regular solution of (B.11) is a
projection of a regular $G$-invariant solution of (B.1) on $K$
determined by (B.10). The last equality in (B.10) defines the
analytical continuation of $\Phi_E(u)$; (B.14) follows from
the second equality in (B.10) and (4.3).

2. Any amplitude, i.e. a scalar product (B.13) of two $S_F$-invariant
states, is independent of a $CS_{ph}$ parametrization (of a gauge choice).
An $S_F$-invariant regular function of $u\in\R^M$ can be decomposed
over the basis $\Phi_E(u)$. Our statement follows from the fact that
theories (B.11-13) corresponding different parametrizations of $CS_{ph}$
are unitary equivalent.

3. Quantization {\em before} eliminating unphysical variables
(the Dirac method) is necessary, {\em otherwise} one might gain a
{\em gauge-dependence} of a quantum theory and, as a result, 1 and
2 do not hold. Indeed, if we would quantize the classical Hamiltonian
(4.18), we encounter the operator ordering problem whose solution
is not unique. However the gauge (parametrization) independence of
the quantum theory can be achieved just at the definite operator
ordering in (B.11-12) uniquely resulting from
the application of the Dirac method as we have seen above.

4. The physical Hamiltonian in (B.11) is $S_F$-invariant
$$
\hat{H}^f_{ph}(u_s(u)) =\hat{H}^f_{ph}(u)\ ,\ \ \ \ \ \ u\in K.
$$
Let us write the Laplace-Beltrami operator $\la\pl/\pl x,\pl/\pl x\ra
=\Delta(\theta, u)$ in the variables (B.4),
push all derivatives $\pl_\theta$ in it to the
right by commuting them with $\theta$ and then set $\pl_\theta=0$.
We denote the operator thus obtained $\Delta_{ph}=\Delta(\theta, u)
\vert_{\pl_\theta=0}$. For any physical state $\Phi=\Phi(u)$, we
have $\Delta(\theta,u)\Phi=\Delta_{ph}\Phi$ because $\hat{\sigma}
\sim \pl_\theta$. Due to the gauge-invariance, the Hamiltonian in (B.1)
commutes with the constraints, $[\hat{H},\hat{\sigma}]=0$ and,
hence, $[\Delta(\theta,u),\hat{\sigma}]=0$ (the potential $V$
is $G$-invariant). Gathering the definition of $\Delta_{ph}$ and
$G$-invariance of $\Delta$ we conclude that $\Delta_{ph}=
\Delta_{ph}(u)$ is independent of $\theta$ (otherwise we would arrive to
the contradiction $0=[\Delta,\hat{\sigma}]\Phi =
\hat{\sigma}\Delta\Phi = \hat{\sigma}\Delta_{ph}\Phi \sim
\pl_\theta\Delta_{ph}\Phi \neq 0$). Consider now the change
of variables $\theta,u\rightarrow \theta_s,u_s$. By its definition
$\Delta(\theta_s,u_s)=\Delta(\theta,u)$ and $\pl_\theta\sim
\pl_{\theta_s}$ (i.e. $\pl_\theta$ does not contain a term
proportional $\pl_{u_s}$ since $\pl u_s/\pl_\theta =0$). This
yields $\Delta_{ph}(u_s)= \Delta(\theta_s,u_s)\vert_{\pl_{\theta_s}=0}
=\Delta(\theta_s,u_s)\vert_{\pl_\theta =0} = \Delta(\theta,u)
\vert_{\pl_{\theta}=0} =\Delta_{ph}(u)$, which completes the
proof of the $S_F-$invariance of the physical Hamiltonian.

To derive a path integral representation of the quantum theory (B.11-13),
we consider a slice approximation of the transition amplitude
$U_t^{ph}(u,u')=\langle u|\exp(-i\hat{H}_{ph}t)|u'\rangle$,
\be
U_t^{ph}(u,u')=\lim\limits_{\epsilon\rightarrow 0}\int\limits_K
\prod\limits_{k=0}^n\left(d^Mu_k\mu(u_k)\right)
U_\epsilon^{ph}(u,u_n)U_\epsilon^{ph}(u_n,u_{n-1})\cdots
U_\epsilon^{ph}(u_1,u')\ ,
\ee
where $(n+1)\epsilon =t$, the limit is taken so that $n\rightarrow \infty,
\ \epsilon\rightarrow 0$, while $t$ is kept fixed; the infinitesimal
evolution operator kernel reads
\be
U_\epsilon^{ph}(u,u')= [1-i\epsilon\hat{H}_{ph}(u)]
\langle u|u'\rangle + O(\epsilon^2)\ .
\ee

A naive limit in (B.15) gives a formal PI with a restricted integration
domain $K\subset \R^M$. A calculation of such a PI meets difficulties
because even a finite dimensional Gaussian integral cannot be explicitly
done over a part of an Euclidean space. In addition, a restriction of
the PI integration domain is meaningless for systems with boundary
conditions appearing  due to a non-trivial topology of a configuration
space like for a particle in a box or on a circle \cite{q}, \cite{kl}
(see also Sec.5). Topological properties of a configuration space
 are taken into account in PI by including additional ``reflected''
trajectories into the sum over pathes \cite{q} rather than by
restricting the PI integration domain. Technically, a relation
between transition amplitudes $U_t$ and $U_t^{eff}$ for the same
systems (the same Hamiltonian) in a topologically non-trivial
$CS$ and in $CS=\R^M$, respectively, is established by means of
an operator $\hat{Q}$ containing all information about a CS topology,
$\hat{U}_t = \hat{U}_t^{eff}\hat{Q}$ \cite{q}. The same form
of PI turns out to be valid in gauge theories \cite{weyl}-\cite{lis}.
Bellow we shall prove this.
For deriving a PI formula we shall use
a method of an analytical continuation of the unit operator kernel
\cite{gr},\cite{book}.

The unit operator kernel $\langle u|u'\rangle$ has a natural
analytical continuation to the unphysical domain $u\in \R^M$.
Indeed, due to the $S_F$-invariance of the basis states (B.14)
we have \cite{gr}-\cite{lis},\cite{book}
\be
\langle u|u'\rangle =\sum\limits_E\Phi_E(u)\Phi^*_E(u') =
\sum\limits_{S_F}\left(\mu(u)\mu(\hat{s}u')\right)^{-1/2}
\delta^M(u-\hat{s}u')\ ,
\ee
where $u\in \R^M,\ u'\in K$. Representing $\delta$-functions
in (B.17) through the Fourier integral and calculating  the
action of $\hat{H}_{ph}(u)$ on the unit operator kernel in
(B.16), we obtain
\ba
U_\epsilon^{ph}(u,u')&=&\int\limits_{\R^M}\frac{d^Mu''}
{(\mu\mu'')^{1/2}}U_\epsilon^{eff}(u,u'')Q(u'',u)\ ,\\
U_\epsilon^{eff}(u,u'')&=&\int\limits_{\R^M}\frac{d^Mp}
{(2\pi)^M}\exp\left[i\epsilon\left(p_j\frac{\Delta_j}{\epsilon}
- H^{eff}(u,p)\right)\right]\ ,\\
H^{eff}(u,p)&=&\frac12 g^{ij}_{ph}(u)p_ip_j + \frac i2
\pl_ig^{ij}_{ph}(u)p_j + V_q(u) +V\ ,\\
Q(u,u'')&=&\sum\limits_{S_F}\delta^M(u-\hat{s}u''),\ \
\ \ u''\in \R^M\ ,\ \ \ u\in K\ ,
\ea
where $\mu''=\mu(u'')$ and $\Delta_j=u_j-u_j''$. So, the
infinitesimal evolution operator kernel (B.18) has the
desired form $\hat{U}_\epsilon^{ph}=\hat{U}_\epsilon^{eff}
\hat{Q}$. A next step is to prove the convolution formula
\ba
U_{2\epsilon}^{ph}(u,u')&=&\int\limits_Kd^Mu_1\mu(u_1)
U_\epsilon^{ph}(u,u_1)U_\epsilon^{ph}(u_1,u')\\
&=&\int\limits_{\R^M}\frac{d^Mu''}{(\mu\mu'')^{1/2}}
U_{2\epsilon}^{eff}(u,u'')Q(u'',u')\ ,\\
U_{2\epsilon}^{eff}(u,u'')&=&\int\limits_{\R^M}d^Mu''
U_\epsilon^{eff}(u,u'')U_\epsilon^{eff}(u'',u')\ ,
\ea
or in the operator form
\be
\hat{U}_{2\epsilon}^{ph}=\hat{U}_\epsilon^{eff}\hat{Q}\hat{U}_\epsilon
^{eff}\hat{Q} = \hat{U}^{eff}_{2\epsilon}\ .
\ee
The proof is given by the following chain of equalities
\ba
U_{2\epsilon}^{ph}(u,u')=\sum\limits_{S_F}\int\limits_K du_1
\frac{\mu_1}{(\mu\mu(\hat{s}u_1))^{1/2}}
U_\epsilon^{eff}(u,\hat{s}u_1)U_\epsilon^{ph}(u_1,u')&=&\\
=\sum\limits_{S_F}\int\limits_Kdu_1|J_s(u_1)|^{1/2}\left(
\frac{\mu_1}{\mu}\right)^{1/2}U_\epsilon^{eff}(u,\hat{s}u_1)
U_\epsilon^{ph}(\hat{s}u_1,u')&=&\\
=\int\limits_{\R^M}
\frac{du''}{(\mu\mu'')^{1/2}}
\sum\limits_{S_F}
\int\limits_Kdu_1|J_s(u_1)|U^{eff}_\epsilon(u,\hat{s}u_1)
U_\epsilon^{eff}(\hat{s}u_1,u'')Q(u'',u')&,&
\ea
where $\hat{s}u_1=u_s(u_1),\ \mu_1=\mu(u_1),\ \mu= \mu(u)$ and
$\mu''= \mu(u'')$. To obtain (B.26), we substitute (B.18) into (B.22) and
do the integral over $u''$. For the transformation (B.27) we
use the relation $\mu(u_s) = \mu(u)/J_s(u)$ (which follows from
the $\tilde{S}_F$-invariance of the measure $d^Nx= (\wedge d\theta)d^Mu
\mu(u) = (\wedge d\theta)d^Mu_s\mu(u_s)$) and the $S_F$-invariance
of the kernel (B.18) (or (B.16)). The latter results from the
obvious relation $\langle\hat{s}u|u'\rangle = \langle u|u'\rangle$
(cf. (B.17)) and $\hat{H}_{ph}^f(\hat{s}u) = \hat{H}_{ph}^f(u)$
(see p.4 above). Equality (B.28) is derived by
substituting (B.18) into (B.27) and using the relation $\mu(\hat{s}u_1)
= \mu(u)/J_s(u)$ again. Finally, (B.28) turns into (B.23) after
changing variables $u_1\rightarrow \hat{s}u_1$ in each term of the
sum in (B.28) by means of the rules (B.8) and (B.7).

For a finite time we get
\ba
\hat{U}_t^{ph}&=& \lim\limits_{\epsilon\rightarrow 0}
\hat{U}_\epsilon^{eff}\hat{Q}\hat{U}_\epsilon^{eff}\hat{Q}\cdots
\hat{U}_\epsilon^{eff}\hat{Q}=\lim\limits_{\epsilon\rightarrow 0}
\hat{U}_\epsilon^{eff}\cdots \hat{U}^{eff}_\epsilon\hat{Q} =
\hat{U}_t^{eff}\hat{Q}\ ,\ \ \ \ \ \\
U_t^{eff}(u,u'')&=&\lim\limits_{\epsilon\rightarrow 0}\!
\int\limits_{\R^M}\!\!\left(\prod\limits_{k=1}^nd^Mu_k\right)
U_\epsilon^{eff}(u,u_n)U_\epsilon^{eff}(u_n,u_{n-1})\cdots
U_\epsilon^{eff}(u_1,u'')\ .
\ea
Formulas (B.29-30) and (B.19) with $\Delta_k/\epsilon = \dot{u}_k
+O(\epsilon)$ solve the problem of the PI construction. Equalities
(B.30) and (B.19) determine a standard slice approximation of PI
over an Euclidean phase space. Removing the slice regularization
in (B.30) we obtain the path integral
\be
U_t^{eff}(u,u'')= \int\limits_{\R^{2M}}
\left(\prod\limits_{\tau = 0}^t\frac{d^Mp(\tau)d^Mu(\tau)}
{(2\pi)^n}\right)\exp i\int\limits_0^td\tau \left(p_j\dot{u}_j
-H^{eff}\right)\ ,
\ee
where the measure implies a sum over all trajectories $u(\tau)$
going from the initial point $u''=u(0)$ to the final one $u=u(t)$.
The physical transition amplitude is given by (B.18) ($\epsilon
\rightarrow t$) and implies a sum over trajectories going from
initial points $u_s(u')= \hat{s}u', \ u'=u(0),\
\hat{s}\in S_F$ rather than from the only one $u'$.
A trajectory going from one of these points, say,
$\hat{s}u'\in K_s,\ \hat{s}\neq 1$, to $u=u(t)\in K$ must cross
the boundary $\pl K$ at a point $\tilde{u}=u(\tilde{\tau})$.
Suppose for simplicity that $u(\tau)\in K$ if $\tau\in (\tilde
{\tau},t)$ and $u(\tau)\in K_s$ if $\tau\in (0,\tilde{\tau})$.
Consider a reflected trajectory composed of two pieces
$\hat{s}^{-1}u(\tau),\ \tau\in (0,\tilde{\tau})$ and $u(\tau),\
\tau\in (\tilde{\tau},t)$, i.e. it connects the initial point
$u'\in K$, the "reflection" point $\tilde{u}\in \pl K$ and the
final point $u\in K$. Due to the $S_F$-invariance of the
effective action, the reflected trajectory gives the same
contribution into the sum over pathes as the "straight" one
$u_s(u')\rightarrow u$. {\em Therefore the PI modification
(B.29) due to a non-trivial topology of $\ph$ (or $CS_{ph}$)
means that in addition to "straight" trajectories $u'\rightarrow u$,
the reflected trajectories $u'\rightarrow \pl K \rightarrow u$
must be included into the sum over pathes}.

\vskip 0.8cm
\renewcommand{\theequation}{C.\arabic{equation}}
\noindent
{\large\bf C.\  Properties of the measure on the gauge orbit space}
\setcounter{equation}0

\vskip 0.8cm
Consider properties of the function (6.12) under transformations from
$W_A$. According to (3.17) we have
\begin{equation}
\kappa (\hat{s}_{\beta ,n}a)=(-1)^{n_\beta n_\rho}\kappa (\hat{s}_\beta a)
\end{equation}
where an integer $n_\rho $ is given by
\begin{eqnarray}
n_\rho &= &\sum\limits_{\alpha >0}^{}\frac{2(\beta ,\alpha)}{(\beta ,\beta
 )} =\frac{4(\beta ,\rho)}{(\beta ,\beta )}\ ,\\
\rho &= & \frac{1}{2}\sum\limits_{\alpha >0}^{}\alpha \ .
\end{eqnarray}
The half-sum (C.3) of all positive roots possesses remarkable properties
\cite{hel}, p.461,
\begin{eqnarray}
&\ &\frac{2(\omega ,\rho )}{(\omega ,\omega )} = 1\ ,\\
&\ &\hat{s}_\omega \rho =\rho -\omega
\end{eqnarray}
for any simple root $\omega$. Since the Weyl group $W$ preserves the root
system and a reflection $\hat{s}_\beta $ in the hyperplane $(\beta ,a)=0$
is a composition of reflections $\hat{s}_\omega $, $\omega$ ranges simple
roots, there exists an element $\hat{s} \in W$ and a simple root $\omega
_\beta $ such that $\hat{s}\omega _\beta =\beta $. Hence, the number
$n_\rho /2$ is an integer,
\begin{equation}
\frac{n_\rho}{2}=\frac{2(\beta ,\rho )}{(\beta ,\beta )}=\frac{2(\omega
_\beta , \hat{s}^{-1}\rho )}{(\omega _\beta ,\omega _\beta )} \in \Z\ .
\end{equation}
Indeed, representing $\hat{s}^{-1} $ as a composition of the generating
elements $\hat{s}_\omega $ and applying (C.4) and (C.5) we
obtain (C.6) since $2(\omega_\beta,\alpha)/(\omega_\beta,\omega_\beta)
\in \Z$ for any root $\alpha$ (Appendix A).
Thus, $n_\rho $ is an even integer, i.e. the factor
$(-1)^{n_\beta n_\rho }$ in (C.1) is equal to 1.

A reflection $\hat{s}_\beta $ permutes roots and therefore for any
positive root $\alpha$ we have $\hat{s}_\beta \alpha =\pm \gamma $
where $\gamma $
is also a positive root, i.e. the function (6.12) may only change its sing
under transformations from the Weyl group. Consider positive roots $\alpha
$ and $\gamma $ such that $\hat{s}_\beta \alpha =-\gamma $. Then
$\hat{s}_\beta \gamma =-\alpha $ because $\hat{s}_\beta ^2=1$, which
implies that a number of positive root changing their sings under the
reflection $\hat{s}_\beta $ is always odd since $\hat{s}_\beta \beta
=-\beta $. This yields
\begin{equation}
\kappa (\hat{s}_\beta a)=-\kappa (a)
\end{equation}
for any root $\beta $.

Since any elements of the affine Weyl group $W_A$ is a composition of the
reflections (3.17), we conclude
\begin{equation}
\kappa (\hat{s}a)=\det \hat{s}\kappa (a) =\pm \kappa (a) ,\ \ \hat{s}\in
W_A\ ,
\end{equation}
where $\det \hat{s} =-1$ if $\hat{s}$ contains an odd number of the
reflections (3.17) and $\det \hat{s}=1$ for the even one.

The function $\kappa (a)$ is an eigenfunction of the Laplace operator in
$\R^r$
\begin{equation}
(\pl _a,\pl _a)\kappa (a)=-\frac{ 4\pi ^2}{a_0^2}(\rho ,\rho )\kappa (a)\
{}.
\end{equation}
A straightforward calculation of the action of the Laplace operator on
$\kappa (a)$ leads to the equality
\begin{equation}
(\pl _a,\pl _a)\kappa (a)=-\frac{4\pi ^2}{a_0^2}(\rho ,\rho )\kappa (a)
+\frac{\pi ^2}{a_0^2} \sum\limits_{\alpha \ne \beta >0}^{}(\alpha ,\beta
)\left [\cot \frac{\pi (a,\alpha )}{a_0}\cot \frac{\pi (a,\alpha
)}{a_0}+1\right]\kappa (a)\ .
\end{equation}
Relation (C.9) follows from
\begin{equation}
\sum\limits_{\alpha \ne \beta >0}^{}(\alpha ,\beta )[\cot (b,\beta ) \cot
(b,\beta ) +1]=\sum\limits_{
\begin{array}{c}
 \small{all}\\
 \small{planes}
\end{array}
}^{}\!\!\! \sum\limits_{
\begin{array}{c}
 \alpha\ne \beta >0\\
 \small{in\  one\  plane}
\end{array}
}^{} (\alpha ,\beta ) [\cot (b,\alpha ) \cot (
b,\beta ) +1]=0\ ,
\end{equation}
where $b\in H$. We have divided the sum
over $\alpha \ne \beta >0$ in (C.11) into sum over two dimentional planes;
each plane contains at least two roots. A root pattern in each plane
coincides with one of the root patterns for algebras of ${\rm rank}\  2,\
su(3),\
sp(4)\sim so(5)$ and $g_2$ because the absolute value of cosine of an
angle between any two roots $\alpha $ and $\beta $ may take only
four values $|\cos \theta _{\alpha \beta }|=\
0,\ 1/\sqrt{2},\ 1/2,\ \sqrt{3}/2$ (see Appendix A). For
the algebras of rank 2, equality (C.11) can be verified by explicit
calculations, i.e. the sum (C.11) for each plane vanishes, which implies
vanishing the whole sum.  In the case of the $su(3)$ algebra, the sum
(C.11) is proportional to
\begin{eqnarray*}
-\cot  b_1\cot  b_2+\cot  b_1 \cot
 (b_1+b_2)+\cot  b_2\cot  (b_1+b_2)+1&= &0\ ,
\end{eqnarray*}
where
$b_{1,2}=(b,\omega _{1,2})$, and $\omega _1, \omega _2$ and $\omega _1
+\omega _2$ constitute all positive roots (see Appendix A). An explicit
form of the quantum potential (6.22) results from (C.9).

The measure (6.12) is proportional to the Weyl determinant \cite{burb},
p.185
\begin{equation}
(2i)^{N_+}\kappa (a)=\prod\limits_{\alpha >0}^{}
\left( e^{i\pi (a,\alpha )/a_0}-e^{-i\pi (a,\alpha
)/a_0}\right)=\sum\limits_{\hat{s}\in W}^{} \det \hat{s}\exp \left(
\frac{2\pi i}{a_0}(\hat{s}\rho ,a)\right)
\end{equation}
with $N_+$ being a number of positive roots, $N_+=(\dim X-r)/2$.

\vskip 0.8cm
\renewcommand{\theequation}{D.\arabic{equation}}
\noindent
{\large\bf D.\  Eigenstates in 2D Yang-Mills theories}
\setcounter{equation}0

\vskip 0.8cm
Substituting $\Phi_n = \kappa^{-1}\Psi_n $ into (6.21) we
find that $\Psi_n$ is an $r$-dimensional plane wave,
$\exp(2\pi i(\gamma,a)/a_0)$. However, not all values of
the momentum vector $\gamma_n\in H$ are admissible. As has
been shown in Appendix B, only regular solutions to (6.21)
have a physical meaning $\footnote{Moreover, singular solutions
do not satisfy the Schroedinger equation in the whole configuration
space (see, for example, \cite{dirq}, pp. 155-156).}$. First of all,
$\Psi_n(a)$ should vanish on the hyperplanes orthogonal to positive
roots, $(\alpha ,a)=0$, because the function $\kappa^{-1}(a)$ has
simple poles on them. Since $(\hat{s}\gamma _n,\hat{s}\gamma _n)=(\gamma
_n,\gamma _n),\ \hat{s}\in W$, the function
\begin{equation}
\Psi _n(a)\sim \sum\limits_{\hat{s}\in W}^{}\det \hat{s} \exp \frac{2\pi
i}{a_0}(\hat{s} \gamma _n,a)
\end{equation}
is an eigenstate of the $r$-dimensional Laplace operator and vanishes as $a$
approaches any of hyperplanes $(a,\alpha )=0$. To prove the latter, let us
decompose $a$ into two parts $a=a^{||}+a^{\perp}$, such that
$(a^\perp,\alpha)=0$ for a root $\alpha $ and let $W^{(\alpha)}$ be the
quotient $W/\Z_2^{(\alpha)},\ \Z_2^{(\alpha)}=\{1,\hat{s}_\alpha\}$,
where $\hat{s}_\alpha \alpha= - \alpha $ and, therefore, $\hat{s}_\alpha
a^\perp = a^\perp,\ \hat{s}_\alpha a^{||} = -a^{||},\ \det\hat{s}_\alpha
= -1$.
Then the sum (D.1) can be rewritten as follows
\begin{eqnarray}
\Psi _n(a) &\sim & \sum\limits_{\hat{s}\in W^{(\alpha )}}^{}\det
\hat{s}\exp \frac{2\pi i}{a_0}(\hat{s}\gamma
_n,a)+\sum\limits_{\hat{s}\in W^{(\alpha)}}^{}\det \hat{s}_\alpha
\hat{s}\exp \frac{2\pi i}{a_0}(\hat{s}_\alpha \hat{s}\gamma
_n,a)= \nonumber \\
&= & \sum\limits_{\hat{s}\in W^{(\alpha )}}^{}\det \hat{s}\exp \frac{2\pi
i}{a_0} (\hat{s}\gamma _n,a)\left[1-\exp \left(-\frac{4\pi
i}{a_0}(\hat{s}\gamma _n,a^{||})\right)\right]\ .
\end{eqnarray}
As $a^{||}$ approaches zero (i.e. $a$ approaches the hyperplane $(a,\alpha
)=0$), the function (D.2) vanishes as $(a^{||},\alpha )$, therefore $\Phi
_n=\Psi _n /\kappa $ has a regular behavior on the hyperplanes $(a,\alpha
)=0$.

In a neighborhood of the hyperplane $(a,\alpha )=n_\alpha a_0,\ \ n\in \Z,\
\ n_\alpha \ne 0,\ \ a^{||}=n_\alpha a_0 \alpha \ \ /(\alpha ,\alpha )+
\epsilon
\alpha $ where $\epsilon \rightarrow 0$. The function(D.2) vanishes as
$\epsilon \rightarrow 0$ if $\ 2(\hat{s}\gamma _n,\alpha )/(\alpha ,\alpha
)$,$\ \ $ $\hat{s}\in W^{(\alpha )}$, is an integer. Since $\hat{s}\alpha
=\beta$
is a root we conclude that the function $\Phi _n=\Psi _n/\kappa $ with
$\Psi _n$ given by (D.1) is regular if
\begin{equation}
\frac{2(\gamma _n,\beta )}{(\beta ,\beta )}\in \Z
\end{equation}
for any root $\beta $.

Eigenvalues $E_n$ in (6.21) read
\begin{equation}
E_n=\frac{\pi \hbar ^2}{a_0^2l}\left[(\gamma _n,\gamma _n)-(\rho ,\rho
)\right]\ .
\end{equation}
For any $\gamma _n$ obeying (D.3),
a vector $\hat{s}_0\gamma _n,\ \hat{s}_0\in W$,
also satisfies (D.3)
and corresponds to the same energy level (D.4) because the Killing form is
$W$-invariant. Exchanging $\gamma _n$ by $\hat{s}_0\gamma _n$ in (D.1) we
have $\Psi _n(a)\rightarrow \det \hat{s}_0\Psi _n(a)$, which means that
linearly independent wave functions corresponding to a given energy level
(D.4) are determined only by $\gamma _n \in K^+$, i.e. $(\omega ,\gamma _n)>
0,\ \omega $ ranges simple roots. Moreover, if $\gamma _n\in \pl K^+$,
meaning that $(\gamma _n,\omega )=0$ for a certain simple root $\omega $,
$\Psi _n(a)=0$. Indeed, changing the summation in (D.1) $\hat{s}
\rightarrow \hat{s}\hat{s}_\omega $ and using $\det \hat{s}_\omega =-1$
with $\hat{s}_\omega \gamma _n=\gamma _n$ (for $(\gamma _n,\omega )=0$) we
get $\Psi _n(a)=-\Psi _n(a)$ and, hence, $\Psi _n(a)=0$.

Due to (D.3) the function (D.1) acquires only the factor $\det
\hat{s}_\alpha $ under transformations (3.17) of its argument. Hence,
regular solutions of (6.21) are invariant with respect of the affine Weyl
group, $\Phi _n(\hat{s}_\alpha a)=\Psi _n(\hat{s}_\alpha a)/\kappa
(\hat{s}_\alpha a)=\Phi _n(a)$ (see (C.8)), which confirms our general
analysis given in Appendix B (cf. (6.23)). Thus, we
do not need to postulate the invariance of physical states with respect to
residual gauge transformation if the Dirac quantization scheme is used.
The boundary conditions (6.23) determining the physical configuration
space topology are automatically fixed in the Dirac method (in contrast with
the reduced phase space quantization where, first, boundary conditions are
imposed by hand and, second, they are not unique \cite{het}, \cite{sem}).

By means of (C.12), we are convinced that regular solutions to (6.21) are
given by the characters $\chi _{\Lambda _n} $ of the irreducible
representations of the gauge group \cite{ch}, p.909
\begin{equation}
\Phi _n(a)=c_n\frac{\sum\limits_{\hat{s}\in W}^{}\det \hat{s}\exp
\frac{2\pi i}{a_0}(\rho +\Lambda _n,\hat{s}a)}{\sum\limits_{\hat{s}\in
W}^{}\det \hat{s}\exp \frac{2\pi i}{a_0}(\rho ,\hat{s}a)}=
c_n\chi _{\Lambda _n}\left(\exp \frac{2\pi ia}{a_0}\right)
\end{equation}
where $c_n$ are normalization constants, $\gamma _n=\rho +\Lambda _n,\ \
\Lambda _n$ labels the irreducible representations (notice that the sum
(D.1) should vanish for all $\gamma _n$ such that $(\gamma _n,\gamma _n
)<(\rho ,\rho )$ because the function (D.3) must be regular, which is
possible
only if $(\gamma _n,\gamma _n )\ge (\rho ,\rho )$). For the character
$\chi _{\Lambda _n}$ we have the following representation (see (6.3))
\begin{equation}
\chi _{\Lambda _n}\left(\exp \frac{2\pi ia}{a_0}\right)=Tr \left(\exp 2\pi
igla\right)_{\Lambda _n}= Tr \left(P\exp ig
\oint\limits_{S^1}^{}Adx\right)_{\Lambda _n}
\end{equation}
where $(e^y )_{\Lambda _n}$ implies a group element $e^y$ in the irreducible
representation $\Lambda _n$. Formula (D.4) shows that solution of (6.1)
and (6.2) are given by the Wilson loops in all irreducible representations
of the gauge group.

The wave functions are orthogonal with respect to the scalar product
(6.14). This follows from the orthogonality of the characters (D.4).
For normalization coefficients $c_n$ we obtain
\begin{equation}
\delta _{nn'}=\int\limits_{K^+_W}^{}da\kappa ^2(a)\Phi _n(a)\Phi ^*_{n'}(a)=
2^{2N_+}c_nc^*_{n'}\int\limits_{K^+_W}^{}da\sum\limits_{\hat{s},\hat{s}'\in
W}^{} \det \hat{s}\hat{s}'\exp \frac{2\pi i}{a_0}(a,\hat{s}\gamma _n
-\hat{s}'\gamma _{n'})\ .
\end{equation}
The integrand in (D.7) is a periodic function on the Weyl cells covering
the Cartan subalgebra, therefore, the integral (D.5) vanishes for all $
\hat{s}\ne \hat{s}'$ because $\gamma _n$ and $\gamma _{n'}$ belong to the
Weyl chamber and the Weyl group acts simply and transitively on the set of
the Weyl chambers. So, there is no Weyl group element $\hat{s}$ such that
$\hat{s}\gamma _n=\gamma _{n'}$ if $\gamma _{n,n'}\in K^+$.
For $\hat{s}=\hat{s}'$ the integral differs from zero
only for $\gamma _n=\gamma _{n'}$ (due to the periodicity of the
integrand). Thus,
\begin{equation}
\vert c_n\vert =2^{-N_+}(N_W\cdot V_{K^+_W})^{-1/2}
\end{equation}
where $N_W$ is a number of elements in the Weyl group, $V_{K^+_W}$ is the
volume of the Weyl cell.

Consider $X=su(2)$. The algebra has one positive root $\omega$. Solutions
to (D.3) are given by $\gamma_n=\omega n/2$ where $n$ ranges positive
integers because $K^+=\R_+$ and $\pl K^+$ coincides with the origin
$\gamma_n=0$. The spectrum and wave functions respectively read
\begin{eqnarray}
E_n &= &\frac{\pi \hbar^2}{4a_0^2l}\ (n^2-1) (\omega,\omega)\ ,\ \ \
n=1,2,...;\\
\Phi_n &= &c_n'\frac{\sin\pi n(a,\omega)/a_o}{\sin\pi(a,\omega)/a_0}\ .
\end{eqnarray}
Substituting $n=2j+1,\ j=0,1/2,1,...$, into (D.9) we observe that
$E_n$ is proportional to eigenvalues of the quadratic Casimir operator
of $su(2)$; $E_n\sim j(j+1)$ where the spin $j$ labels the irreducible
representations of $su(2)$.

Notice, the norm of $\omega$ in (D.9) cannot be chosen arbitrary and
is fixed by structure constants in the
Cartan-Weyl basis, $(\omega,\omega)=1/2$ (Appendix A). If we rescale
roots by a factor, which means, in fact,
rescaling the structure constants in (A.1-3),
the physical Hamiltonian in (6.21) is also changed so that
its spectrum is independent of the rescaling factor.

If one sets $a=a_0\omega\theta/(\omega,\omega)$, then $a\in K_W^+$
implies $\theta\in (0,1)$. The measure $da$ is defined in the orthonormal
basis in $H\sim \R^r$. For the $su(2)$ case we have
$da \equiv da_3,\ a=\sqrt{2}\omega a_3$ so that $(a,a)=a_3^2,\ a_3\in \R$.
Hence, the normalization coefficients $c_n'$ in (D.10) read
\begin{equation}
c_n'=\left(\sqrt{2}a_0\int\limits_0^1d\theta\sin^2\pi n\theta\right)^{-1/2}
=\left(\frac{a_0}{\sqrt{2}}\right)^{-1/2}\ .
\end{equation}

Consider now symmetry properties of eigenstates (D.10) under transformations
generated by homotopically non-trivial gauge group elements (3.4). For
an arbitrary simple compact gauge group, they are determined by shifts
$a\rightarrow a+i/g\Omega_\eta\pl\Omega^{-1}_\eta$ where $\Omega_\eta
=\exp(ix\eta/l)$ so that (cf. (3.12))
\begin{equation}
\exp(2\pi i\eta) = z\in Z_G\ .
\end{equation}
The lattice $\eta$ is given by integral linear combinations of elements
$\alpha/(\alpha,\alpha),\ \alpha$ runs a root systems, because
\begin{equation}
\exp\frac{2\pi i\alpha}{(\alpha,\alpha)}\in Z_G\
\end{equation}
for any root $\alpha$ \cite{hel}, p.311. Let $X$ be $su(2)$. Its only
positive root is $\omega=\tau_3/4$ (see Appendix A). Then $\exp
(2\pi i\omega/(\omega,\omega)) = \exp i\pi \tau_3 = -1 \in \Z_2
=\Z_{su(2)}$. Thus, homotopically non-trivial gauge transformations
are generated by shifts
\begin{equation}
a\rightarrow a+\frac{n\alpha a_0}{(\alpha,\alpha)}\ ,\ \ \ n\in \Z\ .
\end{equation}
Substituting (D.15) into (D.10) we get
\begin{equation}
\Phi_n\left(a+\frac{a_0\omega n}{(\omega,\omega)}\right)
=(-1)^{n+1}\Phi_n(a)\ ,
\end{equation}
i.e. physical states acquire a phase factor under homotopically
non-trivial gauge transformations. The Gauss law (6.2) provides
only the invariance of physical states with respect to gauge
transformations which can be continuously deformed towards the
identity.

\end{document}